\documentclass[preprint]{elsarticle}
\usepackage[utf8]{inputenc}
\usepackage{fullpage}
\usepackage{amsmath}
\DeclareMathOperator{\arctantwo}{arctan2}
\usepackage{amssymb}
\usepackage{graphicx}
\usepackage{overpic}
\usepackage{upgreek}
\usepackage{longtable}
\usepackage{supertabular}
\usepackage{xcolor,soul,framed,caption}
\colorlet{shadecolor}{yellow}
\usepackage{xcolor}
\definecolor{darkred}{rgb}{0.5,0,0}
\definecolor{darkblue}{rgb}{0,0,0.5}
\definecolor{firebrick}{rgb}{0.75,0.125,0.125}
\definecolor{darkgreen}{rgb}{0,0.5,0}
\usepackage[colorlinks=true,linkcolor=firebrick,citecolor=darkgreen,urlcolor=darkblue]{hyperref} 
\usepackage{todonotes}
\usepackage{enumitem}

\usepackage{siunitx}
\usepackage{units}

\usepackage{xspace}
\def\vB{$\vec{v} \times \vec{B}$\xspace}
\def\vvB{$\vec{v} \times  (\vec{v} \times \vec{B})$\xspace}
\def\vBvvB{$\vec{v} \times \vec{B} - \vec{v} \times  (\vec{v} \times \vec{B})$\xspace}
\def\EvB{E_{\vec{v} \times \vec{B}}}
\def\EvvB{E_{\vec{v} \times  (\vec{v} \times \vec{B})}}
\def\fvB{f_{\vec{v} \times \vec{B}}}
\def\fvvB{f_{\vec{v} \times  (\vec{v} \times \vec{B})}}
\def\Egeo{E_\mathrm{geo}}
\def\Ece{E_\mathrm{ce}}
\def\Eem{$E_\mathrm{em}$\xspace}
\def\Erad{$E_\mathrm{rad}$\xspace}
\def\fgeo{f_\mathrm{geo}}
\def\fce{f_\mathrm{ce}}
\def\xmax{$X_\mathrm{max}$\xspace}
\def\dxmax{$D_{X_\mathrm{max}}$\xspace}
\def\dxmaxfit{$D_{X_\mathrm{max}}^\mathrm{fit}$\xspace}
\def\dxmaxtrue{$D_{X_\mathrm{max}}^\mathrm{true}$\xspace}
\def\dxmaxm{D_{X_\mathrm{max}}}
\def\Rgeo{$R_\mathrm{geo}$\xspace}
\def\Rgeom{R_\mathrm{geo}}
\def\sgeo{$\sigma_\mathrm{geo}$\xspace}
\def\sce{$\sigma_\mathrm{ce}$\xspace}
\def\kce{$k$\xspace}

\def\Ageo{$E'_\mathrm{geo}$\xspace}
\def\Ageom{E'_\mathrm{geo}}
\def\Ace{$E'_\mathrm{ce}$\xspace}
\def\Acem{E'_\mathrm{ce}}

\begin{document}
\begin{frontmatter}
\title{An analytic description of the radio emission of air showers based on its emission mechanisms}

\author[1,2]{Christian Glaser\corref{cor1}}
\ead{glaser@physik.rwth-aachen.de}
\author[3,4]{Sijbrand de Jong}
\author[1]{Martin Erdmann}
\author[3,4]{J\"org R. H\"orandel}

\cortext[cor1]{Corresponding author}

\address[1]{RWTH Aachen University, III. Physikalisches Institut A, Aachen, Germany}
\address[2]{Department of Physics and Astronomy, University of California, Irvine, USA}
\address[3]{IMAPP, Radboud University Nijmegen, Nijmegen, Netherlands}
\address[4]{Nikhef, Science Park, Amsterdam, Netherlands}

\begin{abstract}
Ultra-high energy cosmic rays can be measured through the detection of radio-frequency radiation from air showers. The radio-frequency emission originates from deflections of the air-shower particles in the geomagnetic field and from a time-varying negative charge excess in the shower front. The distribution of the radio signal on the ground contains information on crucial cosmic-ray properties, such as energy and mass. A long standing challenge is to access this information experimentally with a sparse grid of antennas. We present a new analytic model of the radio signal distribution that depends only on the definition of the shower axis and on the parameters energy and distance to the emission region. The distance to the emission region has a direct relation to the cosmic ray’s mass. This new analytic model describes the different polarizations of the radiation and therefore allows the use of independently measured signals in different polarization, thereby doubling the amount of information that is available in current radio arrays, compared to what has been used thus far. We show with the use of CoREAS Monte Carlo simulation that fitting the measurements with our model does not result in significant contributions in both systematic bias and in resolution for the extracted parameters energy and distance to emission region, when compared to the expected experimental measurement uncertainties.
This parametrization also enables fast simulation of radio signal patterns for cosmic rays, without the need to simulate the air shower.
\end{abstract}

\end{frontmatter}
\section{Introduction}

Ultra-high energy cosmic rays (UHECRs) impinging onto the atmosphere induce huge cascades of secondary particles. Established techniques for their detection are the measurement of the particles of the air shower that reach the ground, the observation of the isotropic fluorescence light emitted by molecules that have been excited by the shower particles \cite{Auger2014,TA2008} or by non-imaging air-Cherenkov telescopes that measure the incoherent Cherenkov light produced by
the shower particles \cite{TunkaRex2015}. Important observables for most analyses of high-energy cosmic rays are their energy and the atmospheric depth of the shower maximum \xmax, which is an estimator of their mass. In particular the accuracy, i.e., the systematic uncertainty, of the energy measurement is a crucial aspect. The determination of the cosmic-ray energy from stand-alone particle detectors needs to rely on Monte Carlo simulation, where the hadronic interactions have large uncertainties. So far, the best accuracy is achieved with the fluorescence technique, but this is only possible at sites with good atmospheric conditions. Furthermore, precise quantification of the scattering and absorption of fluorescence light under changing atmospheric conditions requires extensive atmospheric monitoring efforts \cite{Auger2014}.

Another independent method for the detection of cosmic rays is the detection of broadband radio-frequency emission from air showers \cite{Huege2016,Schroeder2016}. 
The radio technique combines a duty cycle close to 100\% with an accurate and precise measurement of the cosmic-ray energy \cite{TunkaRex2015, LOFAREnergy, AERAEnergyPRL,AERAEnergyPRD}, as well as a good sensitivity to the mass of the primary cosmic-ray \cite{LOFARNature2016}. In particular, the energy measurement is well-compatible with, and may even outperform, the fluorescence technique in terms of achievable accuracy \cite{PhDGlaser, KrauseICRC2017}. This is mostly due to the transparency of the atmosphere to radio waves and the corresponding insensitivity to changing environmental conditions, and because the radio-frequency emission can be calculated theoretically via first principles from the air-shower development \cite{GlaserErad2016, Gottowik2017}. 

The radio emission from air showers is due to the acceleration and creation of charged particles within the air shower \cite{Endpoint2011} and is described by classical electrodynamics. In practice, particles other than electrons and positrons do not contribute significantly to the radio emission due to their smaller charge-to-mass ratio \cite{Huege2016}. From a macroscopic point of view, radio emission is attributed to two main emission mechanisms: The \emph{geomagnetic} and \emph{charge-excess} emission processes.
In the dominant geomagnetic emission process, electrons and positrons are deflected in the geomagnetic field in opposite directions due to the Lorentz force, resulting in a transverse current. 
The strength of the emission scales with $\sin \upalpha$, where $\upalpha$ is the angle between the particle movement (shower axis) $\vec{v}$ and the geomagnetic field $\vec{B}$.
In the charge-excess emission process, a time-varying negative charge-excess in the shower front leads to a longitudinal current which is mostly due to the knock out of electrons from air molecules.

The spatial distribution of the energy fluence, i.e., the energy per unit area of the radio electric-field pulse, holds information on relevant air shower parameters such as the energy and the atmospheric depth of the shower maximum \xmax \cite{Allan1971a}. 
The amount of energy emitted in the form of radio emission by the air shower -- referred to as the \emph{radiation energy} -- is given by the spatial integral over the energy-fluence. The radiation energy is directly related to the electromagnetic shower energy \Eem and allows for a precise measurement with a theoretical energy resolution of only 3\% \cite{GlaserErad2016}. Thus, the radiation energy serves as a universal estimator of the cosmic-ray energy and is already exploited by the Pierre Auger Collaboration to measure cosmic-ray energies \cite{AERAEnergyPRL, AERAEnergyPRD}.

\begin{figure}[t]
 \centering
 \includegraphics[width=0.35\textwidth]{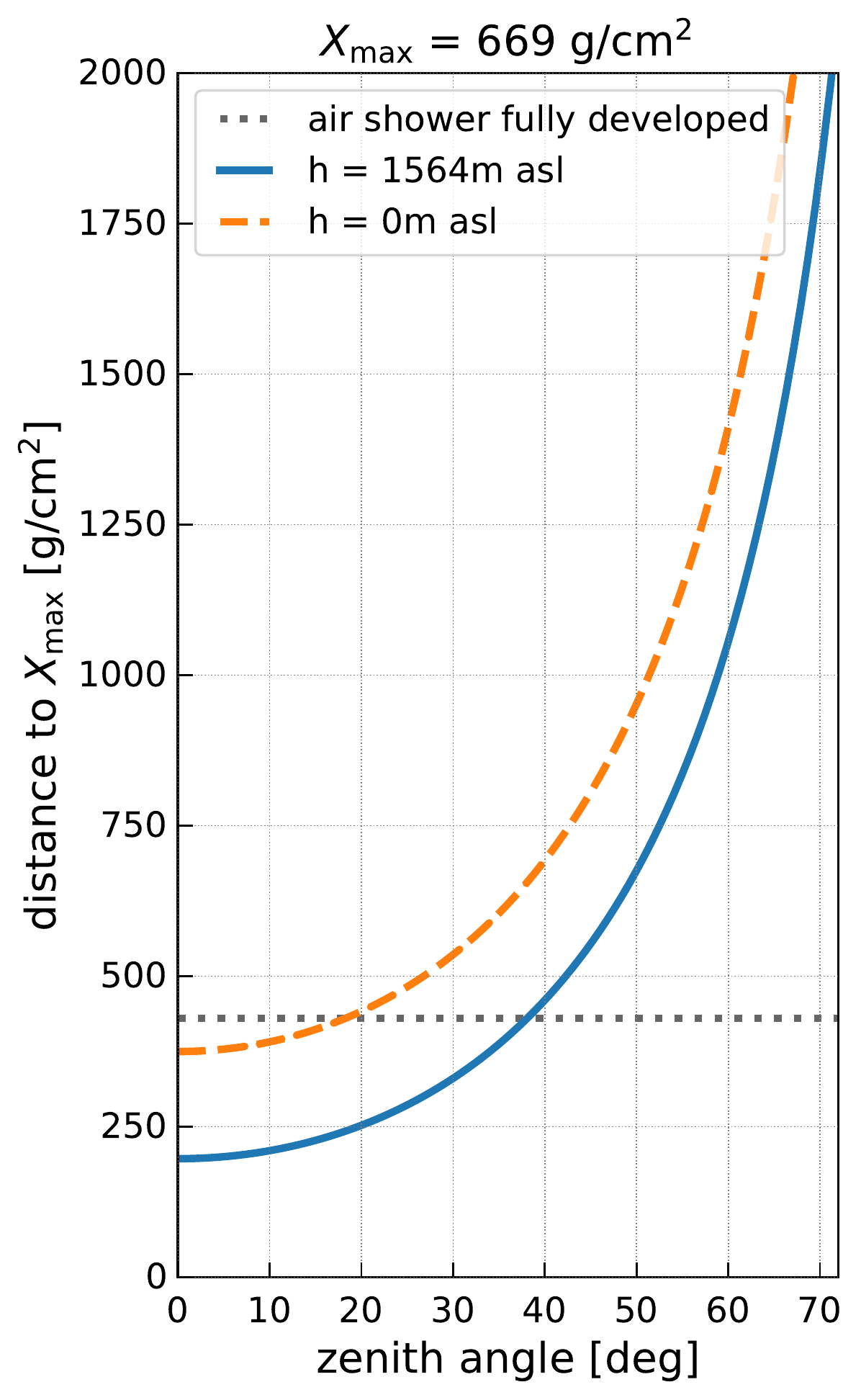}
 \includegraphics[width=0.64\textwidth]{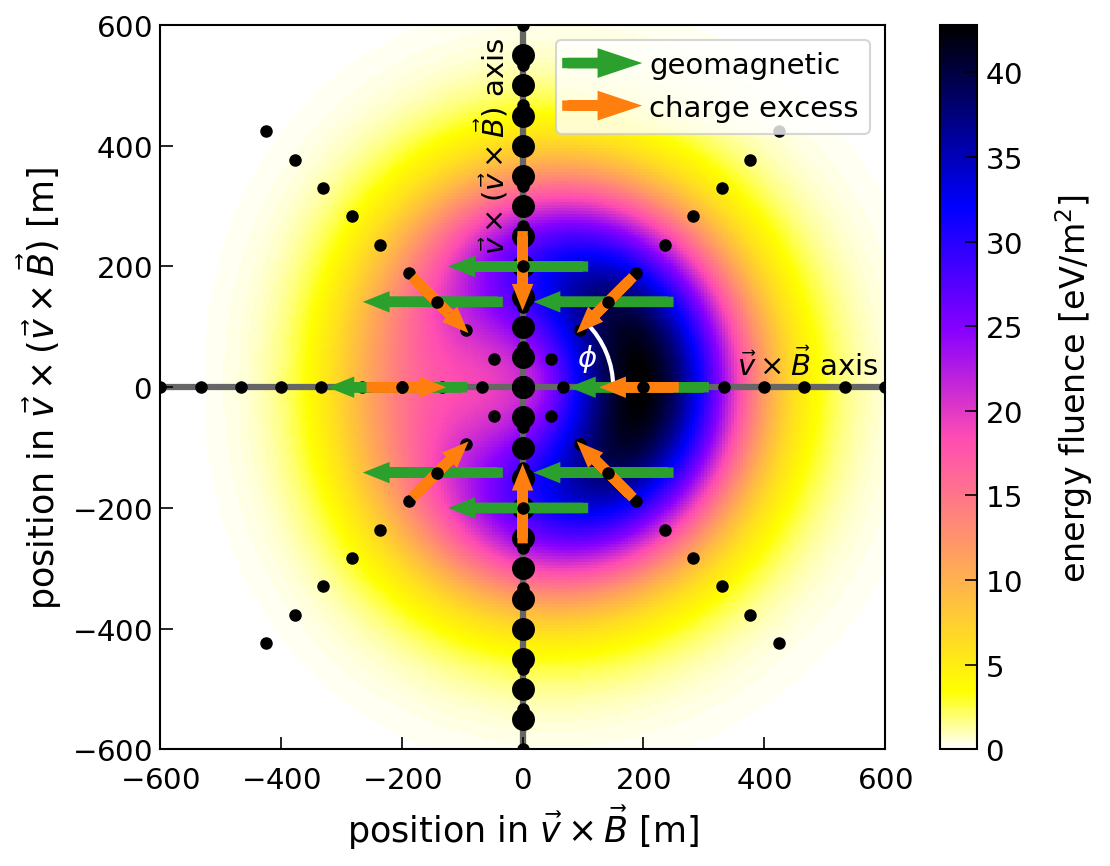}
 \caption{(left) Distance to \xmax as a function of the zenith angle for an average \xmax of \SI{669}{g/cm^2} for two observation altitudes. The dotted line shows the distance to \xmax where the air shower has emitted all its radiation energy. (right) Distribution of the energy fluence (in the \unit[30-80]{MHz} band) of an air shower with 60$^\circ$ zenith angle at an observation altitude of \unit[1564]{m a.s.l.}, which corresponds to the height of the Engineering Radio Array of the Pierre Auger Observatory. Superimposed is the polarization direction of the geomagnetic and charge-excess emission processes at different positions in form of arrows. The black points show the observer positions for which the radio signal was simulated in CoREAS and the larger black points highlight the axis where the signal can be decomposed into the geomagnetic and charge-excess component.}
 \label{fig:starpattern}
\end{figure}

The shape of the spatial signal distribution is primarily determined by the distance \dxmax from the observer to the emission region. The emission region can be approximated by the position of the shower maximum \xmax \cite{GlaserErad2016}. 
The distance \dxmax depends primarily on the zenith angle $\theta$ of the air shower and scales approximately with $\dxmaxm \propto 1/\cos \theta$, with a second order dependence on the value of \xmax for the typical physical range of \xmax \cite{GlaserErad2016}. The dependence is visualized in Fig.~\ref{fig:starpattern} left.
The usage of \dxmax has the advantage that a universal description of the radio signal distribution can be given that does not depend on the specific altitude of the experiment.

A long-standing challenge to access the energy and \xmax information experimentally with a sparse grid of antennas is an analytic modeling of the radio signal distribution and will be addressed in this article. 
In \cite{LOFARLDF}, an empirical parametrization for the spatial radio signal distribution is introduced based on morphological arguments, which gives an adequate description of the data measured by LOFAR and the radio array of the Pierre Auger Observatory (AERA) and was already successfully exploited to measure cosmic-ray energies \cite{AERAEnergyPRL, AERAEnergyPRD}. However, explaining the behavior and value of the parameters of this parametrization is not straightforward, as most parameters depend on various shower features. With the knowledge gained over the past years (e.g. \cite{GlaserErad2016, LOFARLDF, Alvarez-Muniz2014a}), we formulate an analytic description of the spatial signal distribution directly based on its physical emission processes whose parameters directly depend on the air-shower parameters energy, incoming direction and \xmax. In addition, we explicitly use the polarization of the radio signal which effectively doubles the available information of each antenna station. This is achieved by the following approach:

We model the spatial signal distribution on the ground originating from the geomagnetic and the charge-excess emission separately. Then, the two signal-strength distributions are both radially symmetric around the shower axis \cite{GlaserErad2016}. We note that for inclined air showers an additional asymmetry due to the projection of the signal distribution on the ground arises. This imposes no principle problem for our approach but requires an additional correction of the projection effect first. Hence, we restrict our analysis to air showers with zenith angles smaller than 60$^\circ$ where the projection effect is still negligible. Then, the asymmetric two-dimensional radio signal distribution is modeled naturally by the interference of the two emission mechanisms. 
This is because the two emission mechanisms exhibit distinct polarization signatures. The geomagnetic emission is polarized in the direction of the Lorentz force $\vec{v} \times \vec{B}$ acting on the shower particles. The charge-excess emission, in contrast, is polarized radially towards the shower axis. 

The parametrization presented here, also enables the fast simulation of expected signals in a radio detector array. Starting with the species and energy of an incoming cosmic ray and a choice of \xmax and direction of the cosmic ray, the antenna signals can be predicted for any antenna position relative to the shower axis from simple geometric considerations.

In the following, we first present the Monte Carlo data set that we used to develop an analytic description of the geomagnetic and charge-excess function. Then, we present the geomagnetic and charge-excess functions separately and exploit the correlations of the parameters of the functions with the air-shower parameters. Finally, we combine the two functions to model the two-dimensional radio signal distribution. 
Throughout this work we follow the maxim of practical usability of this function, i.e., we demand a precise description of the data with a sufficiently small number of parameters so that it can be applied to current radio air-shower detectors. Following this maxim, we also offer a reference implementation in python that is available on github \cite{geoceLDFgithub}.

\section{Monte Carlo data set and decomposition of radio signal into geomagnetic and charge-excess contributions}

We use the CoREAS program \cite{CoREAS2013}, the radio extension of the CORSIKA code \cite{Corsika}, for the simulation of the radio-frequency emission from air showers. In CoREAS, each shower particle is tracked and the radiation resulting from its movement is calculated from first principles using classical electrodynamics \cite{Endpoint2011}. The radio emission originates only from the movement of electrons and positrons as the contribution from e.g. muons is negligible due to their higher mass-to-charge ratio. This allows for a precise calculation of the radio emission as the development of electromagnetic showers is well understood. 

Recently, many tests have been performed to investigate the accuracy of the radio predictions. On the experimental side, the LOFAR radio cosmic-ray detector with hundreds of antennas with small spacings allows for precise tests of the theoretical predictions. No significant deviation of the CoREAS calculation from experimental data was observed \cite{LofarXmaxMethod, LOFARNature2016}. In addition, a detailed comparison of the CoREAS code with the independent ZHAireS \cite{ZHAireS2012a} air-shower simulation code was carried out and showed no significant difference in the shape of the radio signal distribution \cite{Gottowik2017}.
Hence, we can use the CoREAS code to develop a precise description of the radio signal distribution and study the dependence on air-shower parameters. 

For this analysis we use a set of 300 air showers simulated with CoREAS 7.5602 with QGSJetII-0.4 \cite{QGSJet} and UrQMD \cite{URQMD} as hadronic interaction models. The geomagnetic field is set to an inclination of -35.9$^\circ$ and a strength of \SI{0.24}{Gauss} which corresponds to the value at the Pierre Auger Observatory. We note that this choice does not reduce the general applicability of our results. The scaling of the radio signal with the geomagnetic field is well understood \cite{GlaserErad2016} and our results can be rescaled to different geomagnetic field configurations. The amplitude of the geomagnetic component scales almost linearly with the magnetic field strength whereas the charge-excess component is unaffected by the magnetic field. In \cite{GlaserErad2016} the proper formulas are given to rescale the simulated energy fluences, and we implemented the rescaling in our reference implementation.  The thinning level is set to \num{1e-6} with optimal weight limitation and the lower energy thresholds for electrons/positrons and photons are set to \SI{250}{keV}. 
We use the monthly average atmospheric profile for October at the Pierre Auger site that is available in CORSIKA and corresponds to the yearly average at that site.

A fraction of 50\% of the air showers have an iron primary and 50\% have a proton primary. The cosmic-ray energy is distributed between \SIrange{e17}{e19}{eV}, uniformly in the logarithm of the energy. The zenith angle $\theta$ is distributed uniformly in $\cos \theta$ from $0^\circ$ to $60^\circ$ and the azimuth angle is chosen randomly. For each air shower, we calculate the radio emission for two observations planes, one at \unit[1564]{m\, a.s.l.} -- corresponding to the altitude of the radio array of the Pierre Auger Observatory (AERA) -- and another one at sea level -- corresponding to the altitude of the LOFAR cosmic-ray radio detector.
A suitable coordinate system is in the shower plane (the electric field is always polarized perpendicular to its direction of propagation $\vec{v}$) where one axis is aligned to the \vB direction (the polarization of the geomagnetic component) and the other axis to the \vvB direction. In each observation plane, the observer positions are positioned in a star pattern in this \vB coordinate system projected on the ground plane (see Fig.\ref{fig:starpattern} right). This choice of antenna positions allows for an effective sampling of the two-dimensional radio signal distribution and decomposition of the emission into its geomagnetic and charge-excess contributions (cf. Fig.~\ref{fig:starpattern} right and \cite{GlaserErad2016,LOFARLDF} for more information about the choice of observer positions).
The radio pulses are filtered to be limited to the \unit[30-80]{MHz} band, which corresponds to the bandwidth of most current cosmic-ray radio detectors \cite{AERAEnergyPRD}.

As CoREAS is a microscopic Monte Carlo code, no emission mechanism is explicitly modeled. Therefore, the contribution of the geomagnetic and charge-excess emission processes to the simulated electric field can not be differentiated. However, we can exploit the different polarization signatures of the two emission mechanisms to decompose the signal into its geomagnetic and charge-excess contribution \cite{GlaserErad2016}. 

In Fig.~\ref{fig:starpattern} right, the distribution of the energy fluence is shown in the \vB -- \vvB coordinate system for a typical air shower. At observer positions on the \vvB axis, the polarizations of the signals from the geomagnetic and charge-excess processes are orthogonal. Here, the \vB component of the electric field $\EvB$ originates only from geomagnetic emission, whereas the \vvB component of the electric field $\EvvB$ originates only from charge-excess emission. Hence, we calculate the geomagnetic and charge-excess energy fluences $\fgeo$ and $\fce$ from the respective electric-field components:

\begin{align}
 \fgeo(r) = \fvB(r, \phi = 90^\circ) &= \varepsilon_0 c \Delta t \, \sum\limits_i \EvB^2(r, \phi = 90^\circ, t_i) \label{eq:energyfluence1} \\
 \fce(r) = \fvvB(r, \phi = 90^\circ) &= \varepsilon_0 c \Delta t \, \sum\limits_i \EvvB^2(r, \phi = 90^\circ, t_i) \, ,
 \label{eq:energyfluence2}
\end{align}
where $\varepsilon_0$ is the vacuum permittivity, c is the speed of light in vacuum and $\Delta t$ is the sampling interval of the electric field $\vec{E}(\vec{r}, t)$, which depends on the position $r$, $\phi$ (here in polar coordinates) and time $t$. The positions for $\phi = 90^\circ$ correspond to positions along the positive \vvB axis (cf. Fig.~\ref{fig:starpattern} right).

The shape of the spatial distribution of the energy fluence depends on the distance \dxmax from the observer to the shower maximum. We observed three different categories of shapes corresponding to air showers 
\begin{enumerate}[label=(\Alph*)]
 \item that hit ground before emitting most radiation energy;
  \item that hit ground shortly after emitting all radiation energy; and
 \item that have large distances between the ground and the air-shower development.
\end{enumerate}

In Figs.~\ref{fig:example1} - \ref{fig:example3}, we show typical examples of these three categories. 
The two components of the energy fluence ($\fgeo$ and $\fce$) are presented as a function of the position along the \vvB axis.

\section{Shapes of the signal distribution}

\begin{figure}[t]
 \centering
 \includegraphics[width=0.7\textwidth]{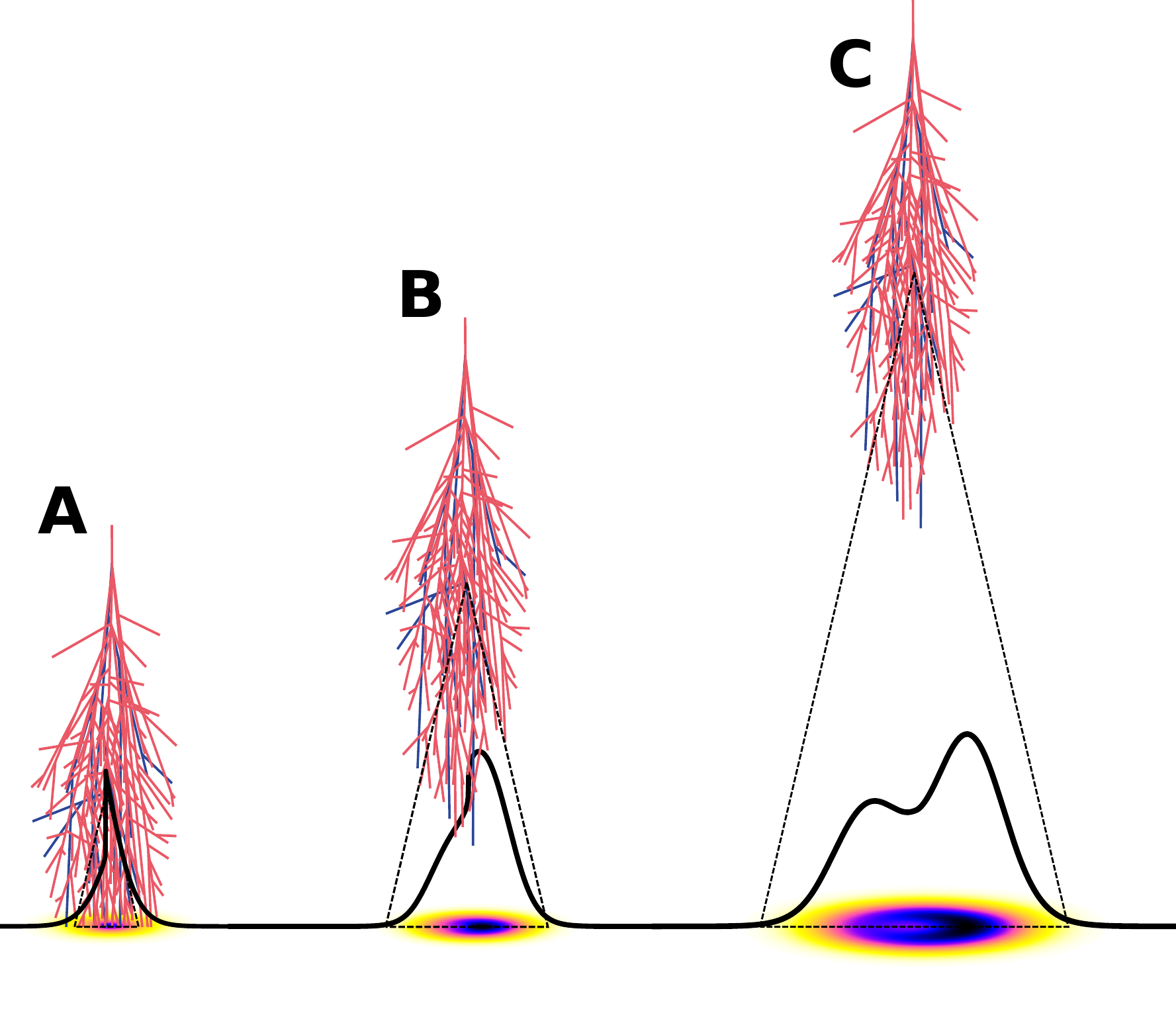}
 \caption{Illustration of how the shape of the distribution of the energy fluence changes with distance to the shower maximum. We note that we only show vertical air showers here for illustration purposes. Throughout the analysis we used air showers with all kinds of incident directions, and the distance to the shower maximum depends strongly on the zenith angle, e.g. example A is a typical shape of a vertical shower whereas example C is a typical example of an inclined shower.}
 \label{fig:sketch}
\end{figure}

\begin{figure}[t]
 \centering
 \includegraphics[width=1\textwidth]{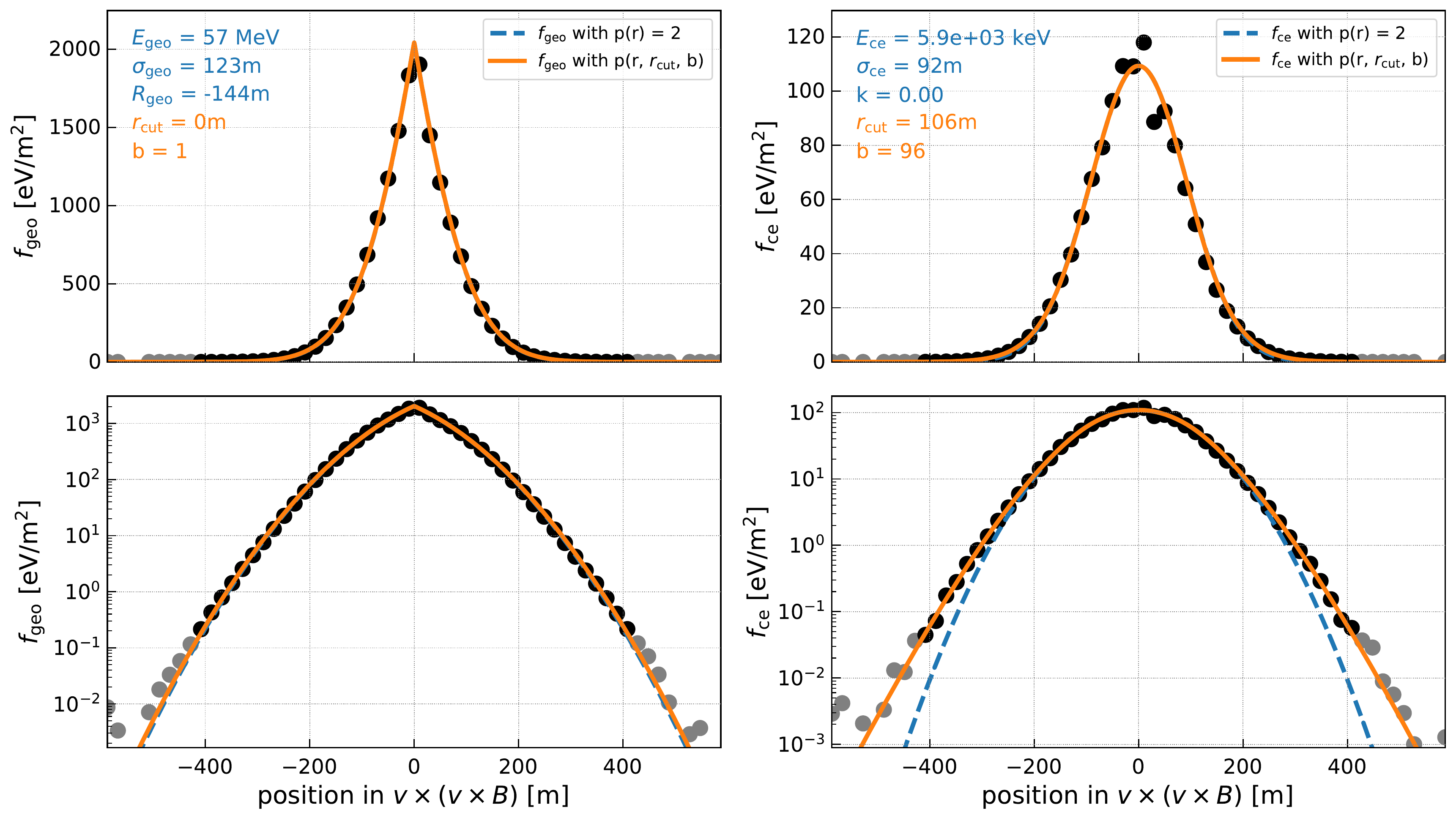}
 \caption{Example A: Energy fluence (in the \unit[30-80]{MHz} band) from geomagnetic (left) and charge-excess emission (right) along the \vvB axis of a \SI{3}{EeV} iron induced air shower with a zenith angle of 32$^\circ$ observed at an altitude of \unit[1564]{m a.s.l.}. The corresponding distance to \xmax is \SI{374}{g/cm^2}. Gray circles denote data points with a signal less than \num{e-4} of the maximum signal that are not used in the fit. The dashed line shows the best fit with $p(r) = 2$ (see below Eq.~\eqref{eq:LDF_geo} and Eq.~\eqref{eq:ce}). The solid line denotes the best fit if the parameters $r_\mathrm{cut}$ and $b$ of $p(r)$ are both varied. The upper panels are on a linear scale and the lower panels are on a logarithmic scale.}
 \label{fig:example1}
\end{figure}

The shape of the spatial radio signal distribution depends on the distance between the observer and the emission region. We call this quantity \dxmax and measure it in grammage along the shower axis. This behavior is illustrated in the sketch of Fig.~\ref{fig:sketch} showing the three typical shapes.
A characteristic number is \dxmax = \SI{430}{g/cm^2} at which the radio emission of the air shower is almost completed (99\% of the radiation energy has already been emitted) \cite{GlaserErad2016}. This means that if a detector measures an air shower at this \dxmax, the shower development will just have completed when hitting the observer. For this or smaller distances to the shower maximum, the distribution of the energy fluence is peaked and narrow around the shower axis (example A). Thereby, the geomagnetic signal distribution is always narrower than the charge-excess signal distribution. E.g., in Fig.~\ref{fig:example1}, the geomagnetic distribution shows a sharp peak at the shower axis, whereas the charge-excess distribution already flattens at the shower axis and shows a Gaussian like shape. 

The second example (B) is for an intermediate distance $\dxmaxm = \SI{572}{g/cm^2}$, where the shower development is already completed but the observer is not yet far away from the emission region. Now, the geomagnetic signal distribution also flattens at the shower axis and its shape is Gaussian like. In the charge-excess case, it starts to become visible that the energy fluence drops to zero at the shower axis. This is an expected behavior as the polarization flips, i.e., changes by 180$^\circ$, at the shower axis. Only if the energy fluence drops towards zero at the shower axis, do we get a continuous transition. We note that also for smaller distances to \xmax the charge-excess energy fluence becomes zero at the shower axis, but on such small scales that it is not visible in the finite sampling of our simulations \cite{GlaserErad2016}.

The third example (C) is for a distance $\dxmaxm = \SI{1046}{g/cm^2}$, far away from the emission region. In particular, the change of the shape of the signal distribution between the second and third example is due to free propagation of the electromagnetic waves and not because additional radio emission is created by the air shower. 
For large distances to \xmax the emission is peaked in a Cherenkov cone, which originates from the refractive index of air being larger than unity. The opening angle of the cone depends on the air pressure at the emission region, i.e., on the height of the emission. The peaking structure of the Cherenkov cone is smeared because the emission occurs in an extended lateral and longitudinal region along the shower axis. 

Independent of \dxmax, we observe that the charge-excess component shows more fluctuations than the geomagnetic component, e.g., in examples B and C, the right wing of the signal distribution shows a slightly higher maximum amplitude than the left wing. For other showers in our data set, both wings have the same maximum amplitude or the left wing has a higher maximum amplitude than the right wing. Similarly, the upward fluctuation near the shower axis of example A (cf. Fig.~\ref{fig:example1} top right) vanishes for other air showers or appears at a different position. Hence, this behavior is likely to be attributed to shower-to-shower fluctuations. Accordingly, our goal in this article is to model the underlying smooth signal distribution and not to model single fluctuations,  although we recognize that once the underlying signal distribution is well modeled these fluctuations may provide interesting additional information on the shower development of an individual event.

\subsection{Discussion of measuring \dxmax in grammage vs. geometric distance}
In this article we measure the distance from the observer to the shower maximum in grammage (\si{g/cm^2})
and not in units of the geometrical distance (\si{km}). This choice is not necessarily obvious because the width of the function should be a function of the geometric distance to the shower maximum if the observer is far away from the emission region. This is because at large distances the shower development has finished
and the radio emission propagates freely through the atmosphere. So one can think of a cone that gets wider
the further away the observer is. However, using the geometric distance to the shower maximum comes with the
following disadvantages:

The transition point, where the shower is fully developed when hitting the surface, has only the same \dxmax for all zenith angles if we measure the distance in \si{g/cm^2}. Hence, we can only describe the transition of the function between the different shapes correctly if we measure \dxmax in \si{g/cm^2}. As a consequence of measuring \dxmax in \si{g/cm^2}, the \dxmax dependence is not completely universal but depends on the model of the atmosphere used in the CORSIKA simulation. Nevertheless, our model will still describe the data/simulations of different atmospheres but the ’\dxmax fit parameter’ has a slight offset to the true \dxmax in the order of
\SI{10}{g/cm^2} - \SI{20}{g/cm^2} (cf. Sec.~\ref{sec:atm} for more details).

The shower development itself depends on \dxmax as a consequence of the air pressure profile: the larger the air density the shorter the geometrical distance in which the shower develops and thereby the smaller the size of the emission region, which in turn influences the spatial distribution of the radio frequency emission. Due to the change in refractivity with air density over the path from the emission region to the detection plane, the radius of the Cherenkov ring also depends on \dxmax, in addition to depending on the geometric distance of the emission region to the detector. Hence, the spatial signal distribution at the detector depends both on \dxmax and on the geometric distance to \xmax. Although these dependencies are not the same, they are rather similar and tend to be degenerate in a fit, except when a huge number of measurements at a large variety of positions are available.

As the goal of this article is to describe the radio signal distribution over the complete \dxmax range, we chose to measure \dxmax in \si{g/cm^2}. For a model of the signal distribution dedicated to horizontal air showers, i.e., only for high zenith angles where the observers are far away from the shower development (case C of Fig. 2), one could use the geometric distance to the shower maximum or a mix of \dxmax and geometric distance. The latter choice would probably also remove the second order dependence of the signal shape on the observation height (cf. Fig.~\ref{fig:dxmax}).

\begin{figure}[t]
 \centering
  \includegraphics[width=1\textwidth]{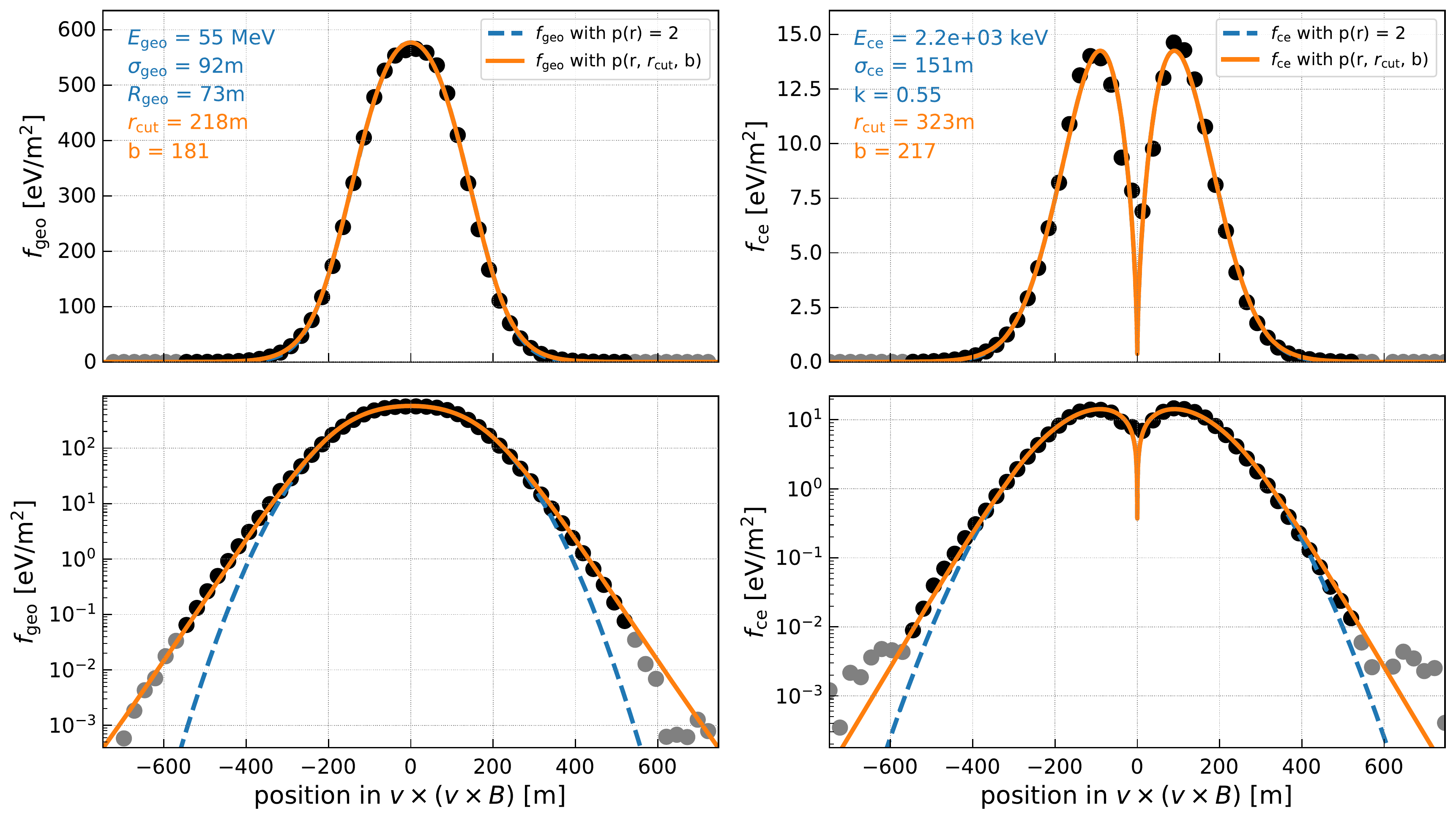}
  \caption{Example B: Same information as in Fig.~\ref{fig:example1} but for a \SI{2.3}{EeV} iron induced air shower with a zenith angle of 46$^\circ$ observed at an altitude of \unit[1564]{m a.s.l.}. The corresponding distance to \xmax is \SI{572}{g/cm^2}. }
  \label{fig:example2}
\end{figure}

\begin{figure}[t]
 \centering
  \includegraphics[width=1\textwidth]{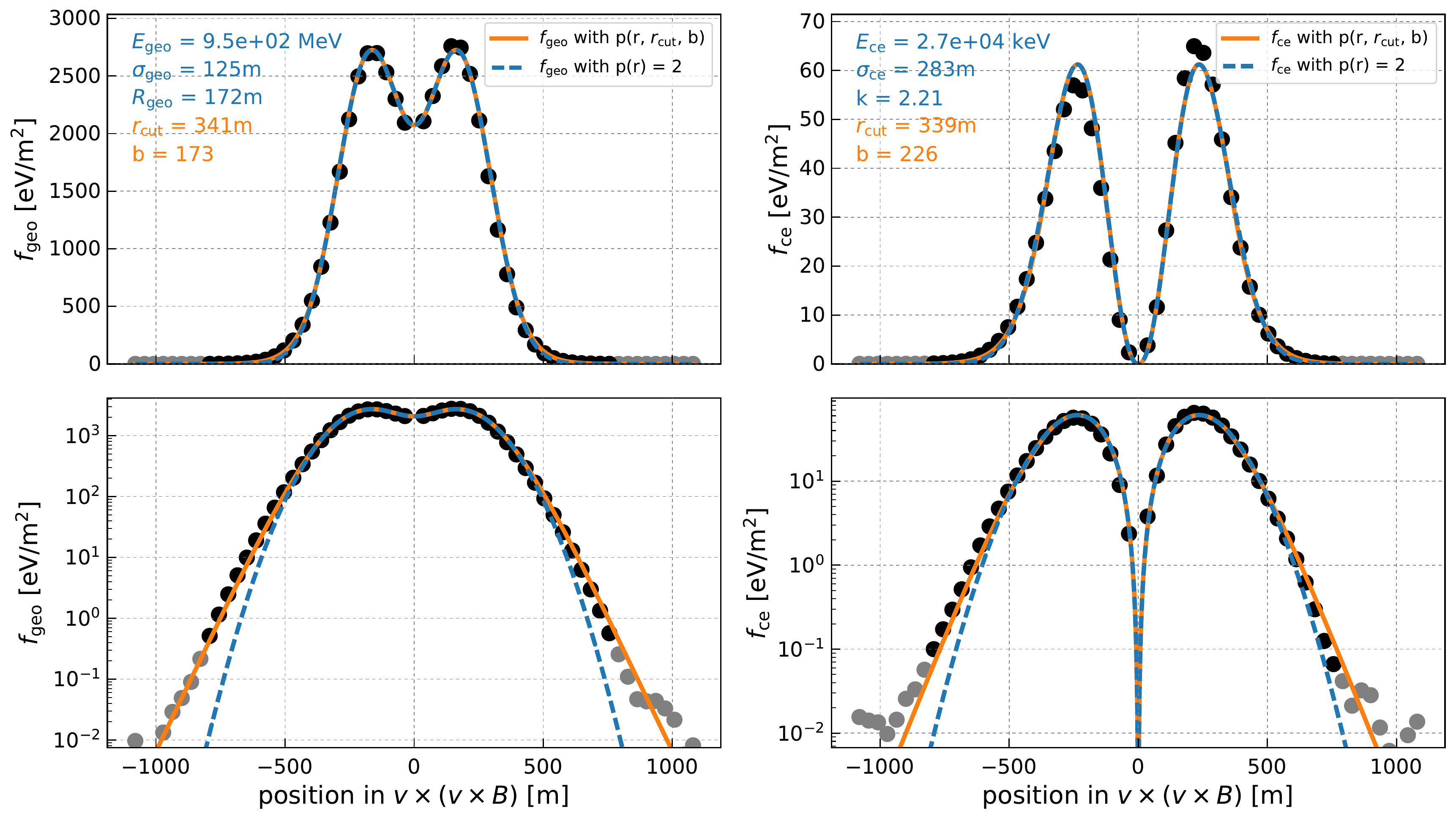}
  \caption{Example C: Same information as in Fig.~\ref{fig:example1} but for a \SI{10}{EeV} iron induced air shower with a zenith angle of 60$^\circ$ observed at an altitude of \unit[1564]{m a.s.l.}. The corresponding distance to \xmax is \SI{1046}{g/cm^2}. The slight asymmetry can be attributed to the projection of the signal at ground onto the shower plane.}
  \label{fig:example3}
\end{figure}

\section{Signal distribution of the geomagnetic emission}

The strength of the geomagnetic emission is circularly symmetric around the shower axis and, thus, only a function of the perpendicular distance to the shower axis $r$. In the \vB coordinate system, $r$ is given by $r = \sqrt{x^2 + y^2}$, where $x$ and $y$ denote the position in the \vB -- \vvB plane. The energy fluence of the geomagnetic emission can be parameterized as
\begin{equation}
 f_\mathrm{geo} = 
\begin{cases}         
\frac{1}{N_{R_-}} \Ageom  \, \, \exp\left(-\left(\frac{r-R_\mathrm{geo}}{\sqrt{2}\sigma_\mathrm{geo}}\right)^{p(r)} \right) &\mbox{if } R_\mathrm{geo} < 0 \\
\frac{1}{N_{R_+}}  \Ageom\left[\exp\left(-\left(\frac{r-R_\mathrm{geo}}{\sqrt{2}\sigma_\mathrm{geo}}\right)^{p(r)} \right) + \exp\left(-\left(\frac{r+R_\mathrm{geo}}{\sqrt{2}\sigma_\mathrm{geo}}\right)^{p(r)} \right)\right] &\mbox{if } R_\mathrm{geo} \geq 0 
\end{cases} \, .
\label{eq:LDF_geo}
\end{equation}
The parameter \Rgeo can be interpreted as the radius of the Cherenkov ring, and the parameter \sgeo describes the width of the function. For \Rgeo $> 0$, the function can be interpreted as signal from a smeared Cherenkov ring that contributes from both sides of the shower axis and thereby fills up the central area in a natural way.
The function $p(r)$ is a small correction to an exponent of $2$ and will be discussed below. For $p(r) = 2$, the two-dimensional integral over the function, which gives the radiation energy, can be calculated analytically. The constants $N_{R_-}$ and $N_{R_+}$ are chosen such that the parameter \Ageo is the geomagnetic radiation energy for $p(r) = 2$. 

\begin{figure}[t]
\def\dx{80}
\def\dy{55}
 \centering
 \begin{overpic}[width=0.32\textwidth]{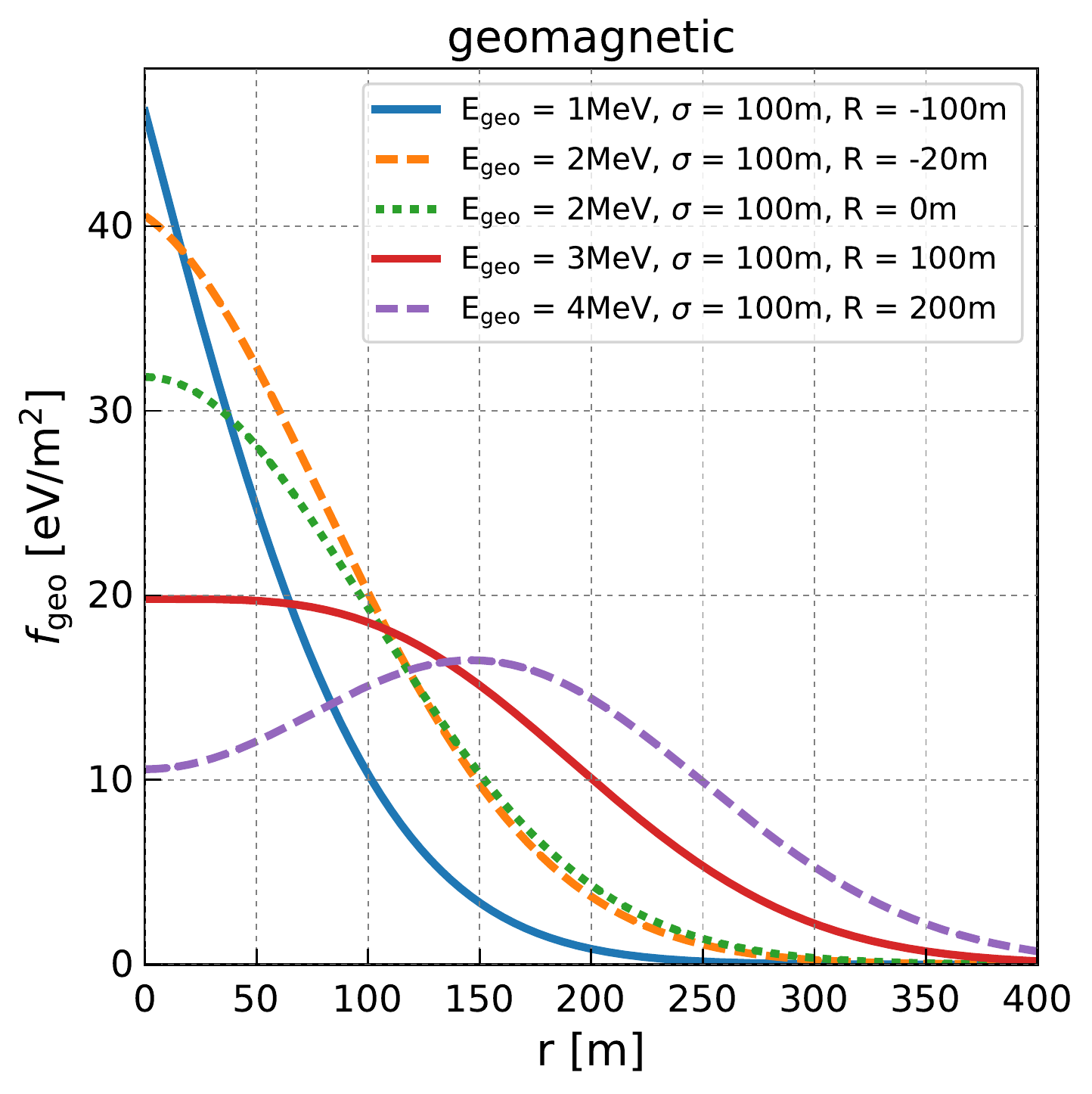}
 \put(\dx, \dy){\Large a)}
 \end{overpic}
 \begin{overpic}[width=0.32\textwidth]{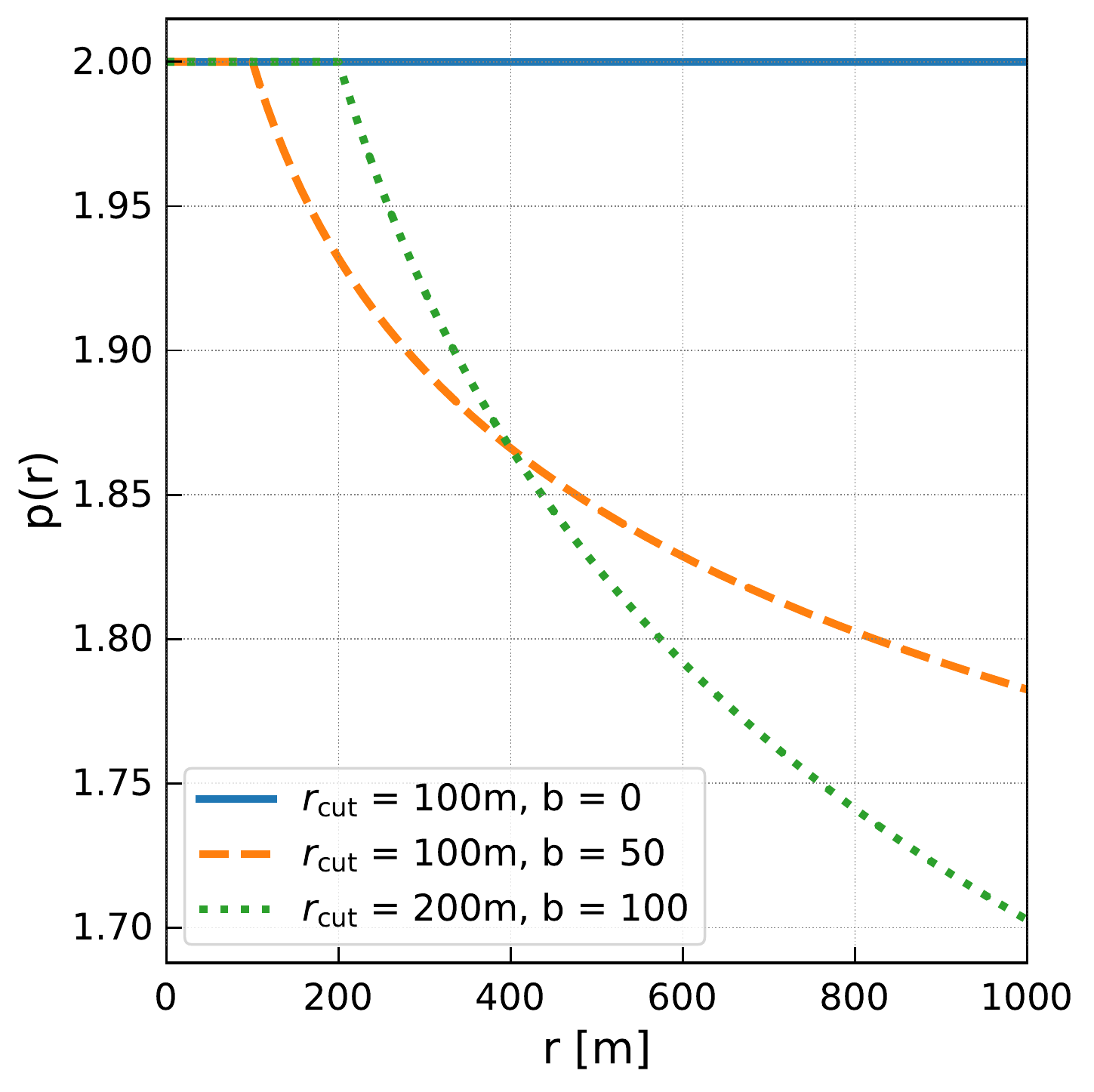}
 \put(\dx, \dy){\Large b)}
 \end{overpic}
 \begin{overpic}[width=0.32\textwidth]{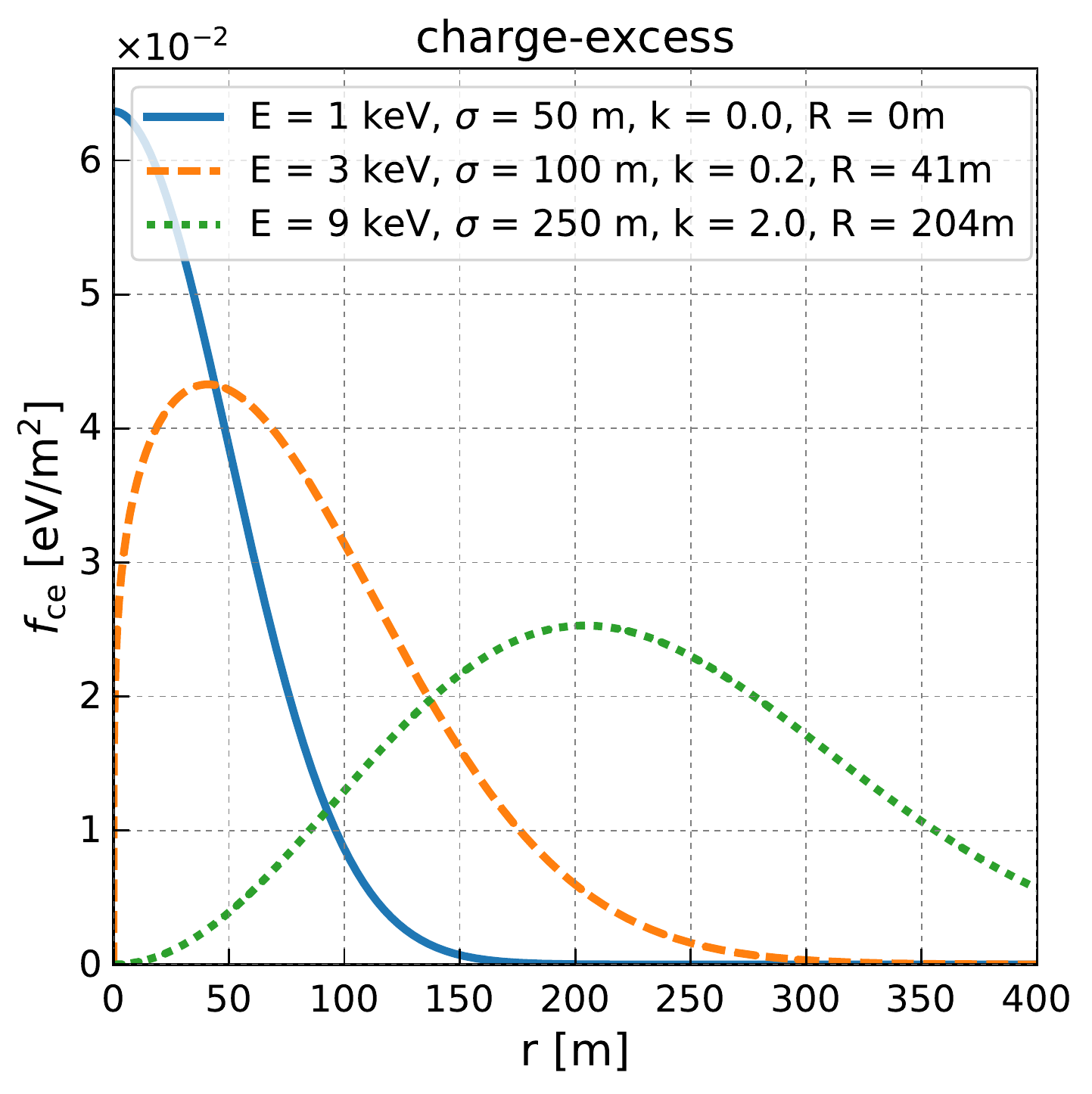}
 \put(\dx, \dy){\Large c)}
 \end{overpic}
  
  \caption{\label{fig:geo_example}(a) Geomagnetic signal distribution for different parameters of our parametrization. (b) Functional form of the variation of the exponent $p(r)$ for different values of $r_\mathrm{cut}$ and $b$. (c)  Charge-excess signal distribution for different parameters of our parametrization.}
\end{figure}

A visualization of the function is shown in Fig.~\ref{fig:geo_example}a.  
Negative values of \Rgeo describe the situation when the air shower has not yet emitted all radiation energy when hitting the observer. Then, the signal distribution is strongly peaked around the shower axis and is described by the falling flanks of a Gaussian function (cf. Fig.~\ref{fig:example1}). Positive values of \Rgeo describe the distribution of the energy fluence after the shower has emitted all its radiation energy (which is roughly at $\dxmaxm \approx \SI{430}{g/cm^2}$ \cite{GlaserErad2016}). Then, the function is the sum of two Gaussian functions centered at \Rgeo and -\Rgeo. If the radius \Rgeo becomes larger then the width \sgeo, the function becomes peaked at the Cherenkov ring. 

\subsection{Determination of optimal parameters}
In this section, the parameters of the geomagnetic function that describe the simulations best are determined in a $\chi^2$ minimization. For each simulated air shower, we do not use all data points in the fit but exclude data points with energy fluences smaller than $10^{-4}$ of the maximum energy fluence where the simulation shows large fluctuations and may be less reliable.  The challenge in this multi-parameter fit is to find a procedure such that the global minimum is found correctly for all signal shapes. We therefore employ the following procedure:

The variation of the exponent is fixed to $p(r) = 2$ in the first iteration.
As the geomagnetic radiation energy can also be calculated via a numerical integration of the data points, we also fix \Ageo in the fit to the result of the numerical integration and determine the optimal parameters $\sigma$ and \Rgeo in a $\chi^2$ minimization. In the fit, the data points with large energy fluence are given a larger weight than those with small energy fluence to have the central part of the function described well. This is done by assigning the same absolute uncertainty to all data points. The fit result can be seen as a dashed blue curve in Figs.~\ref{fig:example1} - \ref{fig:example3}. 

For larger distances of the observer to \xmax (examples B and C where \Rgeo is positive), we observe that the central part of the signal distribution is described well, but the signal fall-off at large distances is slightly overestimated (cf. left panels of Fig.~\ref{fig:example2} and \ref{fig:example3}). This can be modeled by a modification of the exponent of the exponential function of Eq.~\eqref{eq:LDF_geo} of the following form: 
\begin{equation}
p(r) = 
\begin{cases}         
 2 &\mbox{if } r \leq r_\mathrm{cut} \\
 2 \left(\frac{r}{\max(\SI{1}{m},\, r_\mathrm{cut})}\right)^{-b/1000}  &\mbox{if } r > r_\mathrm{cut}
\end{cases}
\quad\,  \, .
\label{eq:p}
\end{equation}
The functional form of $p(r)$ is visualized for typical values of $r_\mathrm{cut}$ and $b$ in Fig.~\ref{fig:geo_example}b. For positive values of $b$, $p(r)$ becomes smaller than $2$ for distances larger than $r_\mathrm{cut}$. Hence, the signal fall-off at large distances weakens. We determine the optimal parameters of $r_\mathrm{cut}$ and $b$ again in a $\chi^2$ minimization, where we fix all other parameters (\Ageo, \sgeo and \Rgeo). We give all data points the same relative uncertainty to increase the weight of the data points with small energy fluence. With this modification the simulated signal distribution can be described well at all distances $r$. The geomagnetic function with the modification of the exponent is shown as a solid orange curve in the left panels of Fig.~\ref{fig:example2} and \ref{fig:example3}.

For small distances to \xmax, where the signal distribution is modeled with negative values of the parameter \Rgeo, a variation of the exponent $p(r)$ is not necessarily required. However, we observe that the parameters \Rgeo and \sgeo are not very well constrained. Larger negative values of \Rgeo can be compensated by larger values of \sgeo such that the $\chi^2$ function has no clear minimum. 
While this imposes no problem for the reconstruction of the radiation energy, the correlation between \Rgeo and \dxmax is disturbed. 
We found that allowing for a variation of the exponent $p(r)$ solves this problem. As the parameters $r_\mathrm{cut}$ and $b$ are correlated with \sgeo and \Rgeo, a separate fit as in the $\Rgeom > 0$ case will not work. Instead, we first determine \Ageo, \sgeo and \Rgeo with $p(r) = 2$ as described above with the additional constraint of $\Rgeom > \SI{-200}{m}$. Then, we fit all five parameters of the function simultaneously with the start parameters of \Ageo, \sgeo and \Rgeo set to the values of the previous fit result and without any constrains on \Rgeo. In the combined fit we assign the same relative uncertainty to all data points to increase the weight of the data points with small signal strength. 

\subsection{Dependence of fit parameters on the distance to the shower maximum}
The goal of this section is to reduce the number of fit parameters by exploiting correlations with the distance to the shower maximum. In the end, the geomagnetic function should depend only on the radiation energy, which is the integral of the function, and \dxmax which determines its shape. 
In the following, we first parametrize the variation of the exponent $p(r, r_\mathrm{cut}, b)$ (Eq.~\eqref{eq:p}) as it is only a small correction to the shape of the function. Then, we repeat the fit with $p(r, r_\mathrm{cut}, b) = p(r, \dxmaxm)$ fixed to its \dxmax parametrization (note that \dxmax is a known quantity in this simulation study). This will result in a more stable determination of the optimal parameters of \sgeo and \Rgeo and less fluctuations in their correlation with \dxmax. 

\begin{figure}[tp]
\def\dx{33}
\def\dy{100}
 \centering
 \begin{overpic}[width=0.4\textwidth]{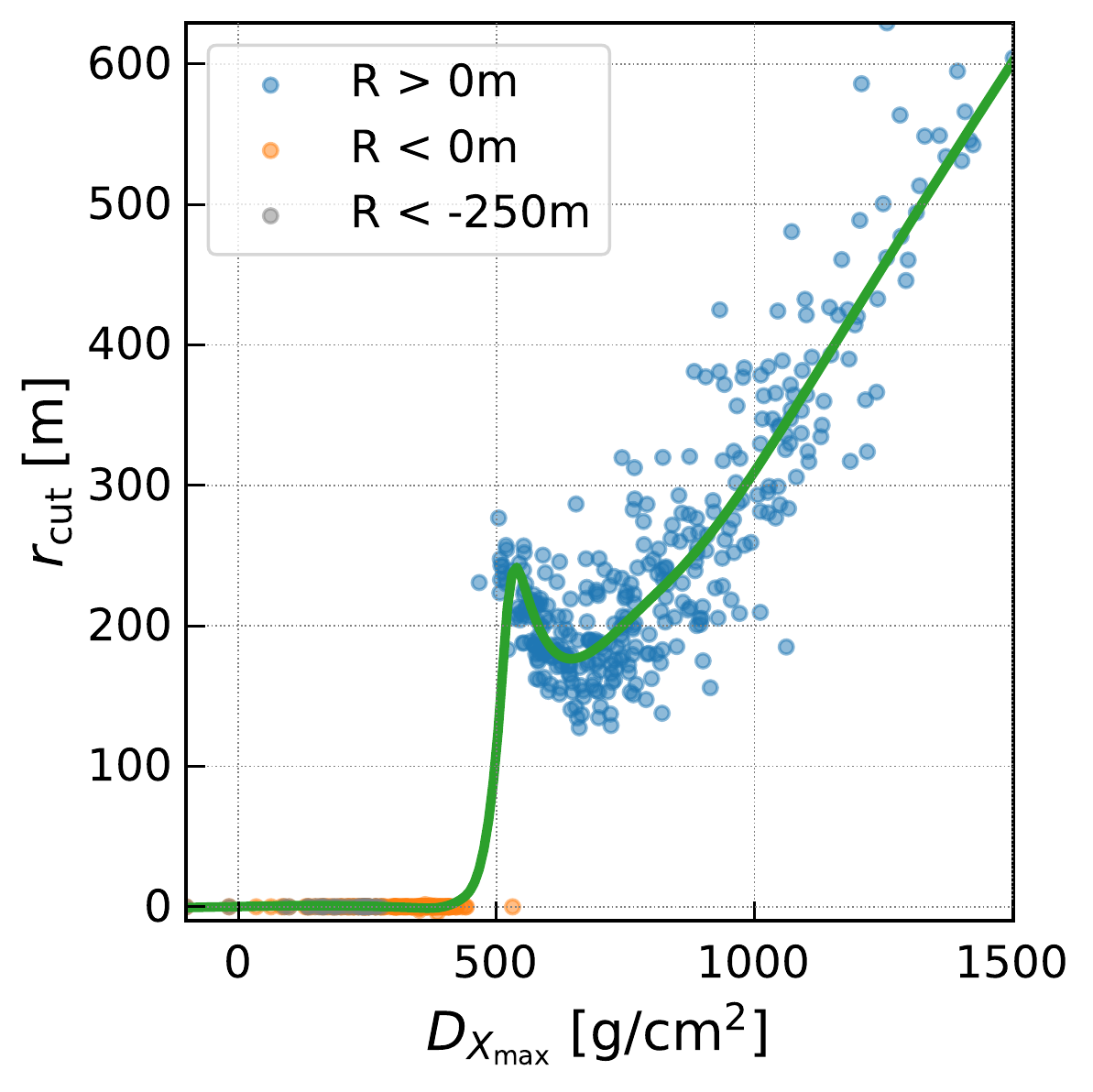}
 \put(\dx, \dy){\Large geomagnetic}
 \end{overpic} 
  \begin{overpic}[width=0.4\textwidth]{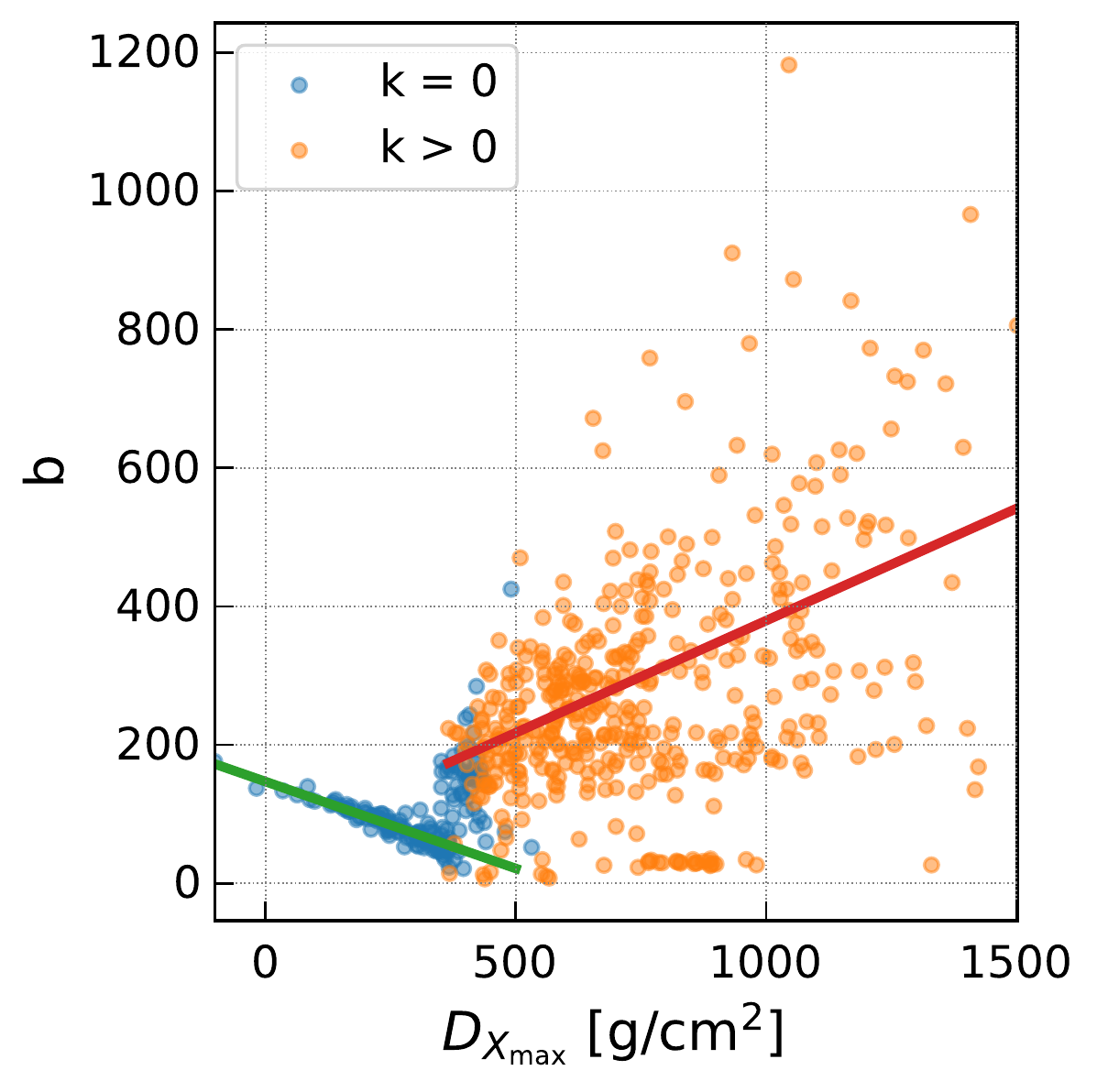}
 \put(\dx, \dy){\Large charge excess}
 \end{overpic} 

  \includegraphics[width=.4\textwidth]{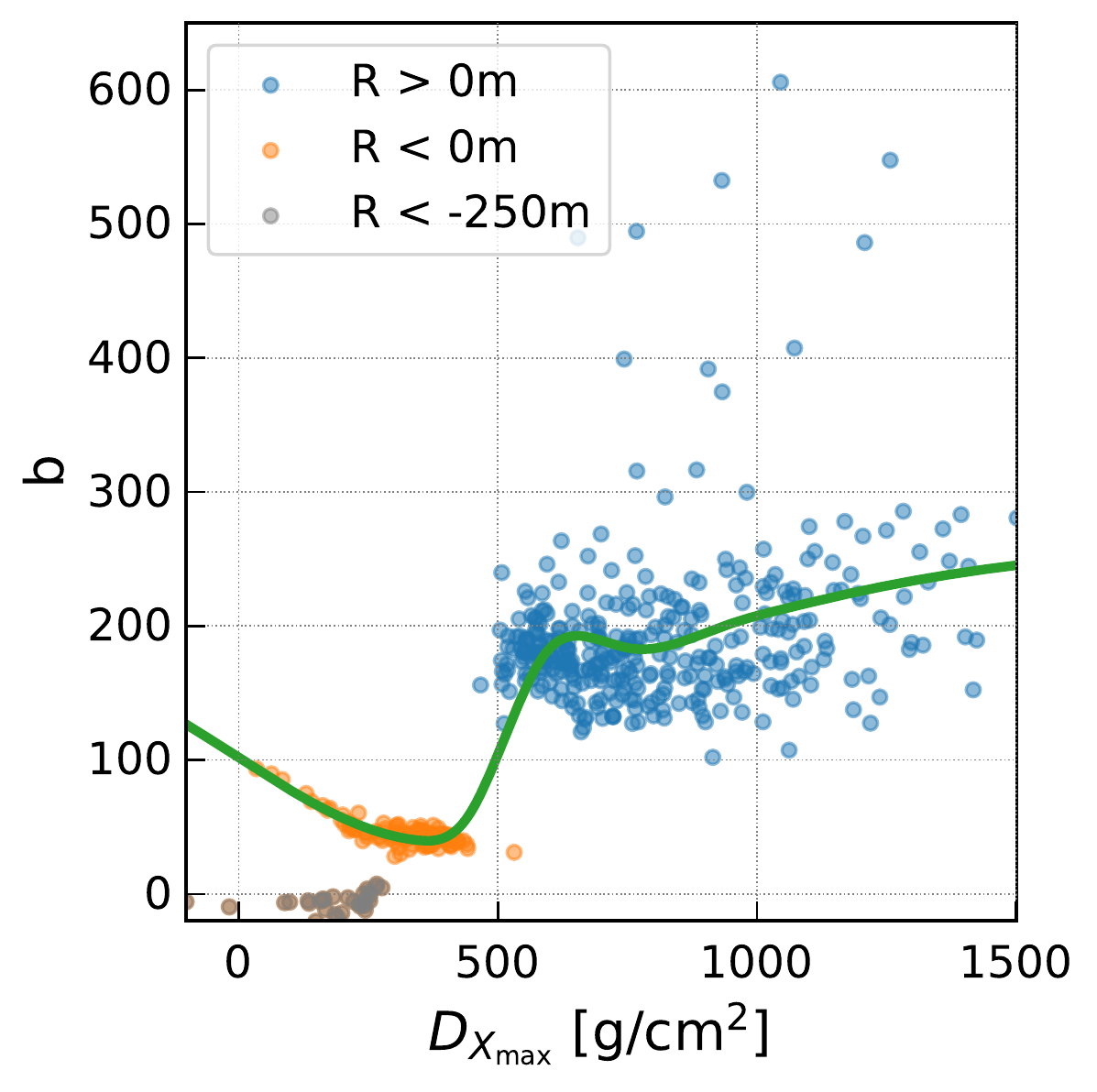}
  \includegraphics[width=.4\textwidth]{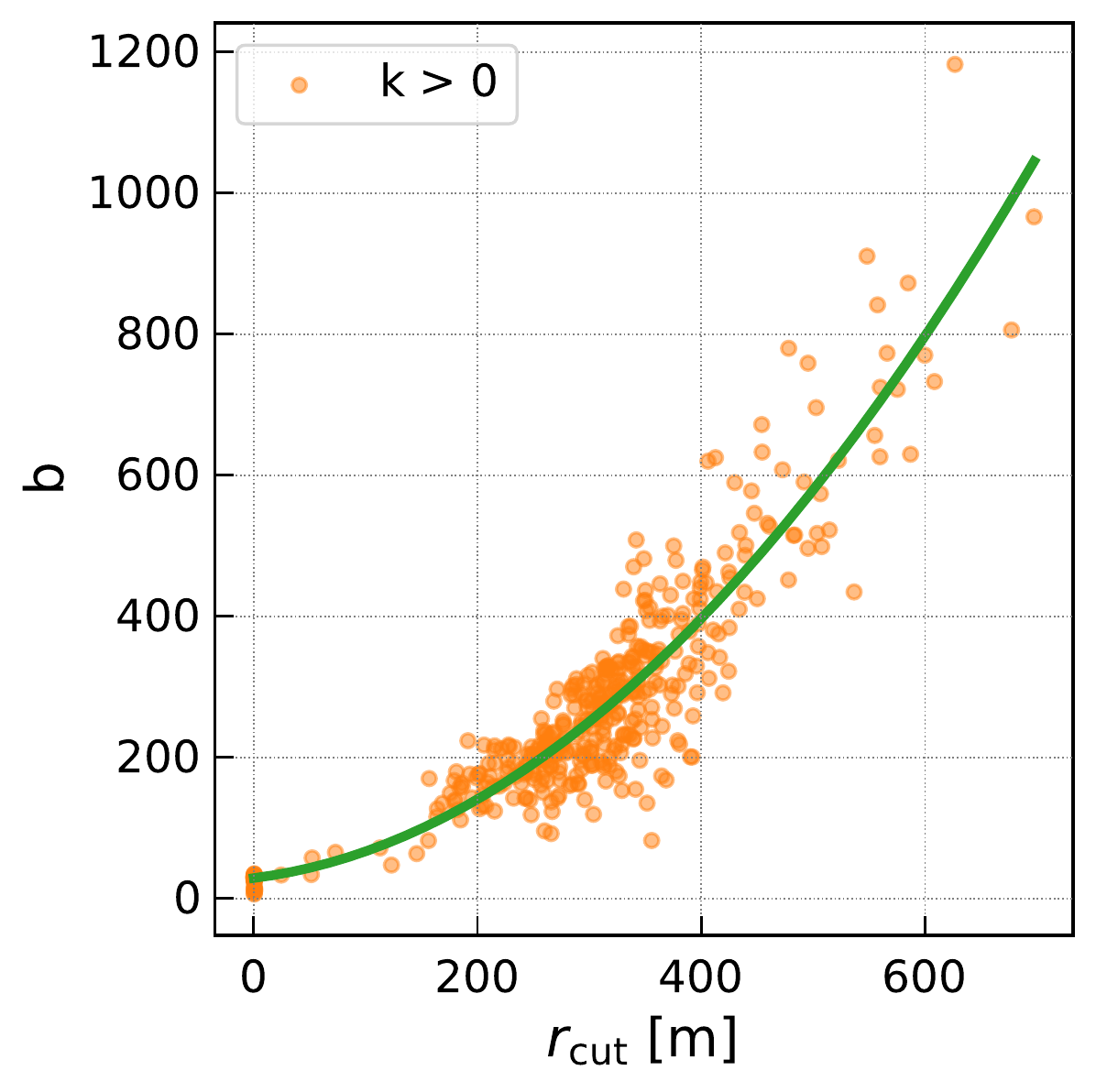}
  \caption{Dependence of the parameters $r_\mathrm{cut}$ and $b$ of Eq.~\eqref{eq:p} on \dxmax. The solid line shows the analytic parametrization. (left) geomagnetic: The data points are subdivided into fully developed air showers where the parameter $R$ is positive and not fully-developed air showers where the parameter $R$ is negative. 
  For $R < 0$ the fit is sometimes unstable and leads to $R$ parameters much below \SI{-250}{m} that are neglected in this parametrization. The solid curves show the parametrization with B-spline functions. (right) charge-excess: Again, the data set is divided in fully and not fully developed showers ($k > 0$ and $k = 0$). In the upper right plot, the two straight lines show the parametrization for $k > 0$ and $k = 0$. The lower right plot shows the correlation of $b$ with $r_\mathrm{cut}$. For $k = 0$, the parameter $r_\mathrm{cut}$ is essentially zero and not shown here. For $k > 0$, the correlation between $b$ and $r_\mathrm{cut}$ is parameterized with a second-degree polynomial.}
  \label{fig:pdxmax}
\end{figure}

\paragraph{Parametrization of $p(r)$}
In the left panels of Fig.~\ref{fig:pdxmax}, the correlations of $r_\mathrm{cut}$ and $b$ with \dxmax are shown for all air showers in our data set. The correlation shows a different behavior for functions with $\Rgeom > 0$ and with $\Rgeom < 0$. In addition, some fits converged at large negative values of \Rgeo, which result in $b$ parameters close to zero. Therefore, we ignore all data points with $\Rgeom < \SI{-250}{m}$ in the parametrization of $b(\dxmaxm)$.\footnote{After fixing $p(r)$, the parameter \Rgeo will not show this behavior anymore and remains within reasonable limits.}

The correlation of $r_\mathrm{cut}$ and $b$ with \dxmax are both parameterized with spline functions. Spline functions have the advantage of being capable of describing arbitrary relations analytically with a small set of parameters. The technical details of using spline functions are discussed in \ref{sec:spline}. In how much detail a spline function describes the data depends on its number of parameters. This \emph{smoothness} is controlled by an external parameter during the determination of the optimal spline function. 
We adjust the smoothing condition manually such that the main trend of the correlation is followed but the function does not follow a single fluctuation. In particular, we adjusted the smoothing condition such that the parametrization has a smooth transition between the $\Rgeom < 0$ and $\Rgeom > 0$ cases. 
The resulting spline function is shown as green curve in Fig.~\ref{fig:pdxmax}.  
The parameters of the function are tabulated in our reference implementation \cite{geoceLDFgithub}.

\begin{figure}[t]
 \centering
  \includegraphics[width=.4\textwidth]{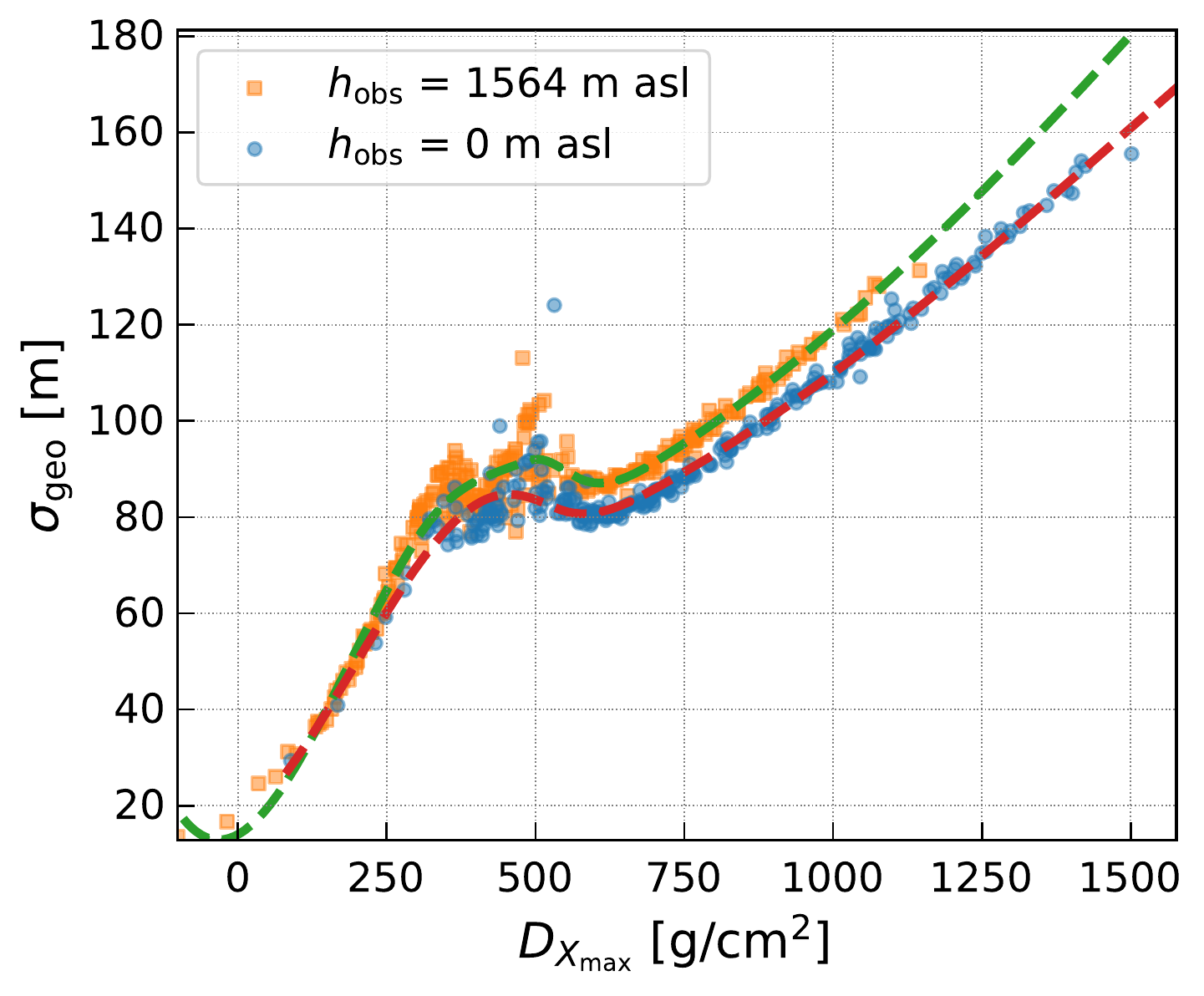}
  \includegraphics[width=.4\textwidth]{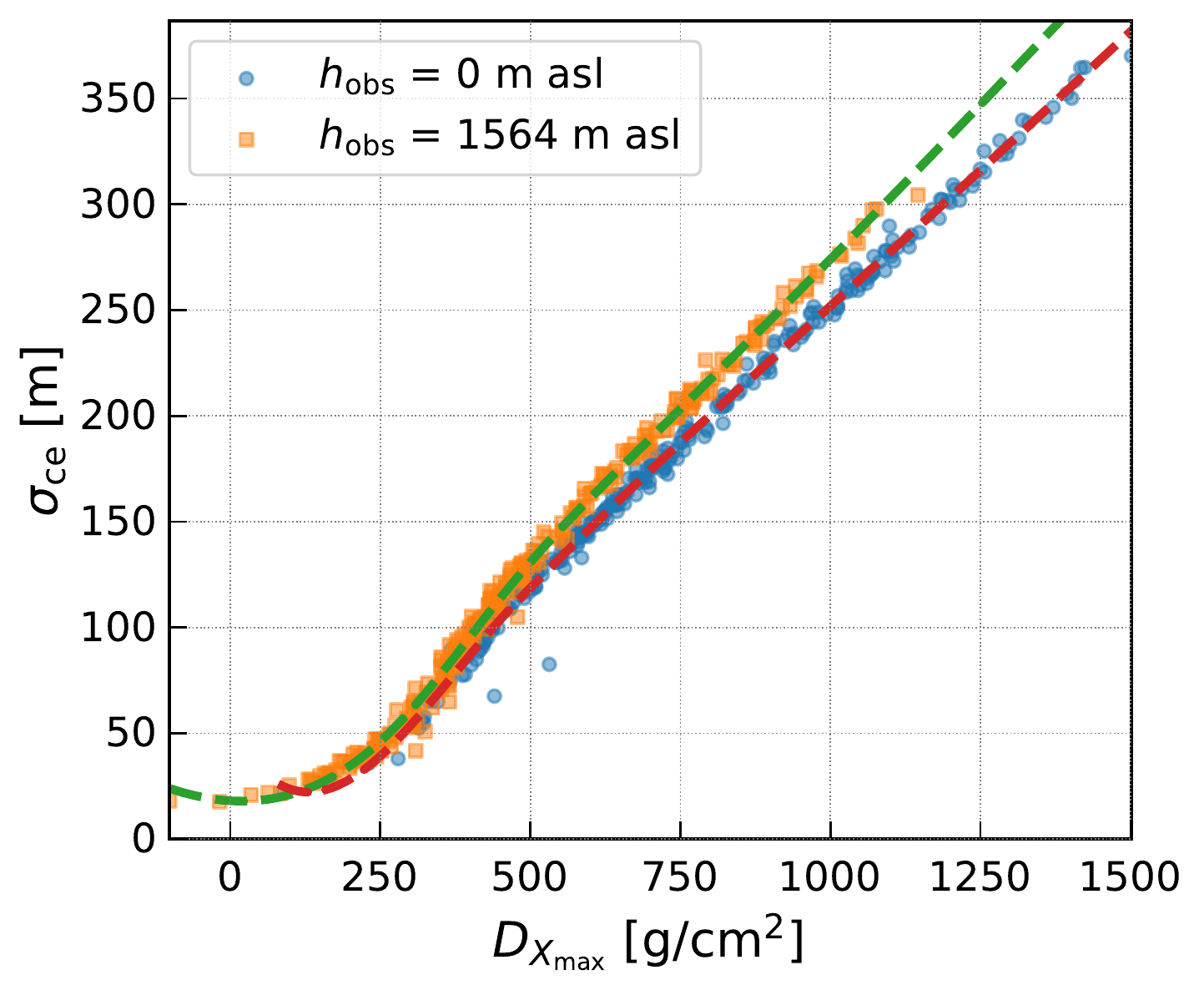}
  \includegraphics[width=.4\textwidth]{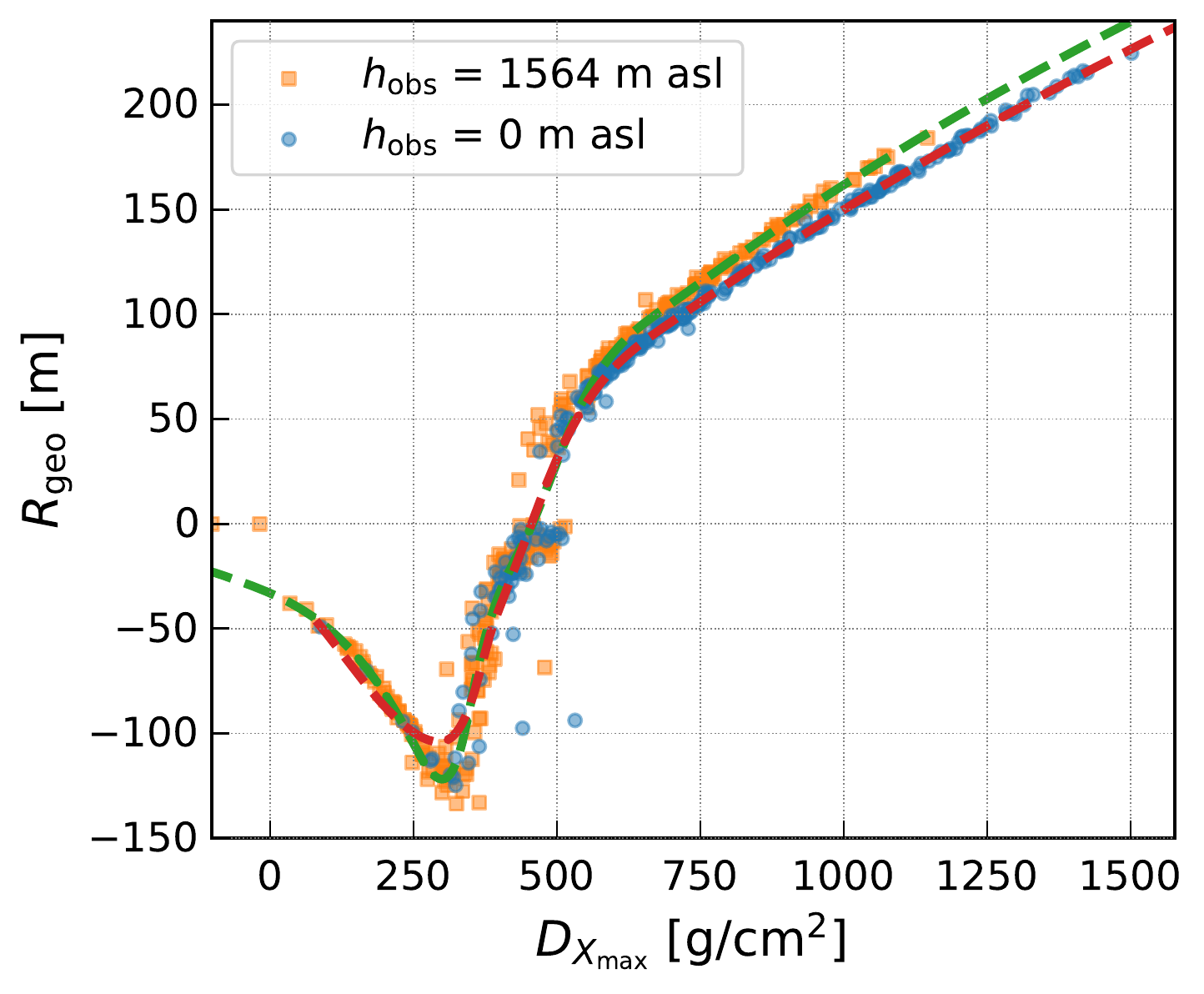}
  \includegraphics[width=.4\textwidth]{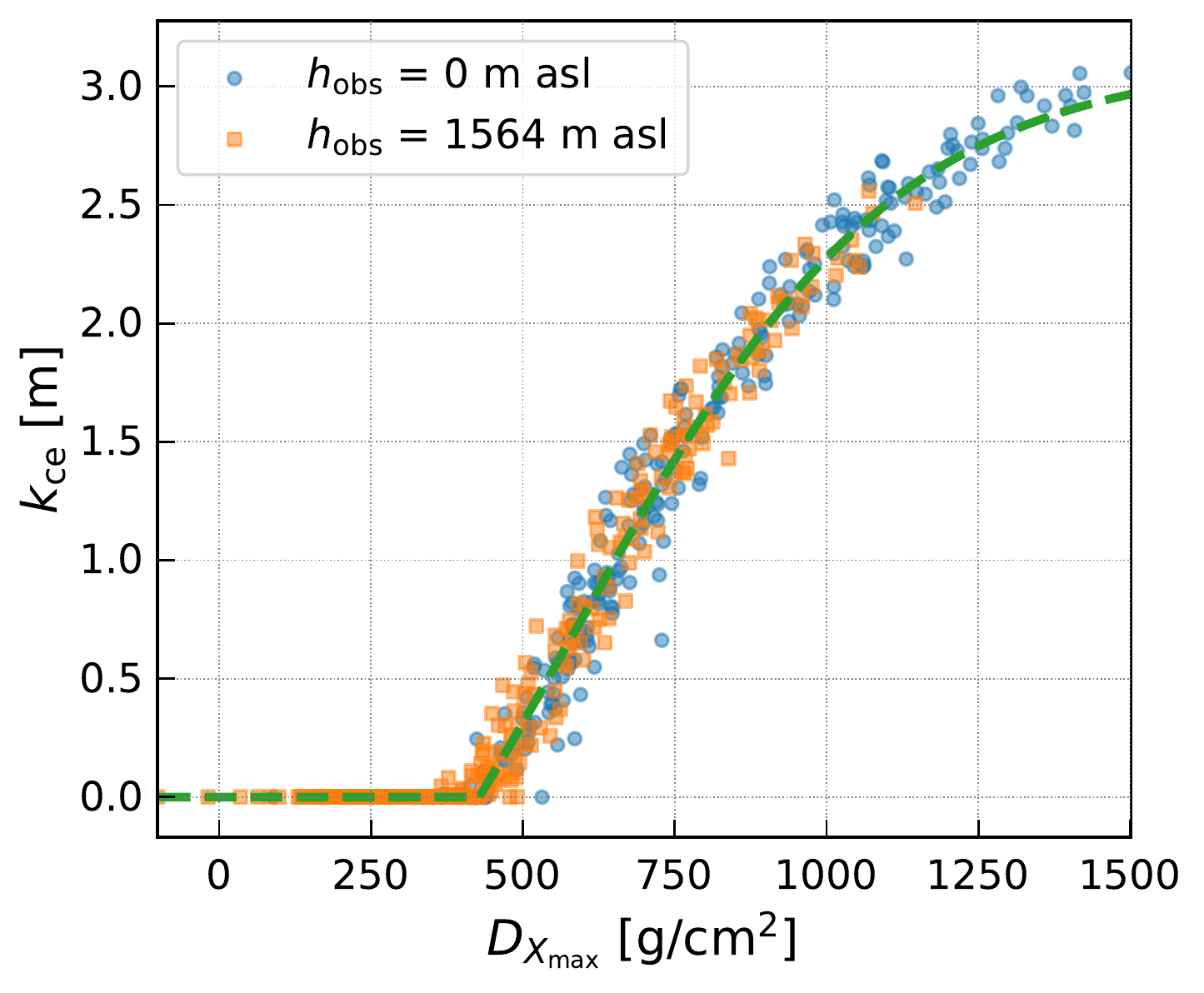}
  \caption{Dependence of fit parameters of the geomagnetic function (left) and the charge-excess function (right) on \dxmax. The dashed lines show the analytic parameterizations of the correlation using B-spline functions. This particular combination of $\sigma_\mathrm{geo}(\dxmaxm)$ and $R_\mathrm{geo}(\dxmaxm)$ results in a geomagnetic function $f(\dxmaxm$) that transitions smoothly between the shapes of different $\dxmaxm$ regions. }
  \label{fig:dxmax}
\end{figure}

\paragraph{Parametrization of \sgeo and \Rgeo}
In the next step, the geomagnetic function (Eq.~\eqref{eq:LDF_geo}) is fitted again to our data set with $p(r)$ fixed to its \dxmax parametrization. Then, the correlation of \sgeo and \Rgeo with \dxmax can be studied, which is presented in the left panels of Fig.~\ref{fig:dxmax}. We observe that the correlation is slightly different for observers at different altitudes. Hence, we perform a separate parametrization for an observation altitude of \SI{1564}{m\, a.s.l.} (the height of the Pierre Auger Observatory) and at sea level (the height of the LOFAR detector). Again, the flexibility of spline functions allows us to parameterize the correlations in a continuous and smooth way. Their parameters are presented in \cite{geoceLDFgithub}. In case of fully developed air showers ($\dxmaxm \gtrsim \SI{430}{g/cm^2}$), both \sgeo and \Rgeo show a smooth, nearly linear increase with \dxmax. The parameter \Rgeo increases faster than \sgeo. At around \SI{600}{g/cm^2}, \Rgeo becomes larger than \sgeo resulting in a visible Cherenkov ring.

For smaller \dxmax, the dependence is more complex and difficult to interpret due to the interplay between \sgeo and \Rgeo. In particular, it is difficult to model the transition from a Gaussian shaped to a narrowly peaked signal distribution, i.e., the transition from fully developed showers to showers that are still developing when hitting the observer. 
Although the individual dependencies $\sigma_\mathrm{geo}(\dxmaxm)$ and $R_\mathrm{geo}(\dxmaxm)$ are not monotonous, their combination leads to a geomagnetic function $f_\mathrm{geo}(\dxmaxm)$ that is smooth in \dxmax. This is, with increasing \dxmax, $f_\mathrm{geo}$ shows a smooth transition from a narrowly peaked distribution, via a Gaussian shaped distribution, to a broad distribution with a visible Cherenkov ring. Thus, it fulfills the primary objective of this article: An analytic description of the radio signal distribution whose shape is determined by one variable only, the distance to \xmax. As this behavior is not directly obvious from Fig.~\ref{fig:dxmax}, we provide a video of the development of the geomagnetic function with \dxmax as supplemental material \cite{videogeoce}.

We now managed to parameterize the geomagnetic function in terms of only two air-shower parameters: \Ageo and \dxmax. The only issue that still needs our attention is that \Ageo does not correspond directly to the radiation energy $\Egeo$ because of the variation of the exponent $p(r)$. This is because Eq.~\eqref{eq:LDF_geo} can be integrated analytically only for $p(r) = 2$ and we normalized the function only for the $p(r) = 2$ case (see \ref{sec:normalization}). However, we can integrate Eq.~\eqref{eq:LDF_geo} numerically for each value of \dxmax and parameterize the deviation between $\Egeo$ and \Ageo as a function of \dxmax. Again, we use splines to parameterize this relation and present the parameters in \cite{geoceLDFgithub}. The final function is presented later in the text in Eq.~\eqref{eq:LDF_geo2}.

\section{Signal distribution of the charge-excess emission}

The strength of the charge-excess emission is circular symmetric around the shower axis and can be described with a modification of the Gamma distribution
\begin{equation}
 f_\mathrm{ce}(r) = \frac{1}{N_\mathrm{ce}} \, \Acem \, r^k \exp\left(\frac{-r^{p(r)} (k+1)}{p(r) \sigma_\mathrm{ce}^{p(r)}}\right) \, ,
 \label{eq:ce}
\end{equation}
with $k \geq 0$. The variation of the exponent $p(r)$ has the same functional form as in the geomagnetic case (cf. Eq.~\eqref{eq:p}). For $k = 0$ (and $p(r) = 2$), the function is a Gaussian function with mean zero. For $k > 0$ the function has the property to be zero at $r = 0$. The interplay between the rising part from $r^k$ and the falling part from the exponential function models the Cherenkov ring. The distance where the function becomes maximal is given by $R_\mathrm{ce} = \sigma_{ce} \sqrt{k} / \sqrt{k + 1}$. The constant $N_\mathrm{ce}$ is chosen such that the two-dimensional integral over $f_\mathrm{ce}$ is $\Ece$ for $p(r) = 2$. Hence for $p(r) = 2$, \Ace equals the radiation energy of the charge-excess emission $\Ece$.  A visualization of $\fce$ is shown in Fig.~\ref{fig:geo_example}c. 

For small distances to \xmax, the signal distribution is maximal and peaked at the shower axis, which is described with $k = 0$ (cf. Fig.~\ref{fig:example1}). We note that for $k=0$ the energy fluence does not become zero at the shower axis. Here, the electric-field vector changes its sign and the energy fluence should become zero. However, as discussed above, the drop towards zero at the shower axis occurs at such small scales that it is not detectable in any experiment as the typical size of an antenna is larger than the distance at which the energy fluence becomes zero. In particular, the sampling of our simulations is not fine enough to see this effect at small distances to \xmax.  Also in practice, not modeling this effect in our function does not have any effect on the radiation energy (the integral over the function) nor on its dependence on \dxmax.

For larger distances to \xmax, it becomes visible that the energy fluence goes to zero at the shower axis. Hence, \kce becomes larger than zero to model the observed behavior (cf. Fig.~\ref{fig:example2} and Fig.~\ref{fig:example3}). We find that a modification of the exponent $p(r)$ leads to better results at large distances to the shower axis for all distances to \xmax.

\subsection{Determination of optimal parameters}
To obtain the optimal fit parameters, we follow the same procedure as for the geomagnetic case. We first determine the parameters \sce and \kce in a $\chi^2$ minimization where the radiation energy is fixed to the result of a numerical integration of the data points, $p(r)$ is fixed to $2$, and all data points are given the same absolute uncertainty to increase the influence of the data points with high energy fluence. The resulting charge-excess functions are shown as blue dashed curves in Figs.~\ref{fig:example1} - \ref{fig:example3}.

Then, in a separate fit, the optimal parameters $r_\mathrm{cut}$ and $b$ of the variation of the exponent are determined. In this fit, \Ace, \sce and \kce are fixed to the previous fit results and the same relative uncertainties are given to all data points. %

\subsection{Dependence of fit parameters on the distance to the shower maximum}
As in the geomagnetic case, we first parameterize the relation of $b$ and $r_\mathrm{cut}$ with \dxmax, which is shown in the right panels of Fig.~\ref{fig:pdxmax}. We observe a different behavior for $k > 0$ and $k = 0$. For $k = 0$, the parameter $r_\mathrm{cut}$ is always zero and the dependence of $b$ on \dxmax can be described with a straight line. For $k > 0$, the relation between $b$ and \dxmax is also described by a straight line. However this time, $r_\mathrm{cut}$ is not zero but shows a correlation with $b$ that can be described by a second order polynomial. From this correlation we can calculate the dependence of $r_\mathrm{cut}$ on \dxmax
The parameters of the parametrization are presented in \cite{geoceLDFgithub} and in \ref{sec:parametrizations}.

In the next step, the charge-excess function is fitted again to our data set with $p(r)$ fixed to its \dxmax parametrization. Then, the correlation of \sce and \kce with \dxmax can be studied, which is presented in Fig.~\ref{fig:dxmax}. In the case of \sce, we again observe that the correlation is slightly different for observers at different altitudes. Hence, the correlation is parameterized separately for \SI{1564}{m\, a.s.l.} and sea level. We model the correlation with B-spline functions and present their parameters in \cite{geoceLDFgithub}. 

For the parameter \kce we do not observe any difference between different observation altitudes. Before the air shower has emitted all its radiation energy (at $\dxmaxm \approx \SI{430}{g/cm^2}$), \kce is zero. For larger distances \dxmax, \kce increases monotonously with \dxmax. The relation can be described well with a logistic function of the form 
\begin{equation}
 k_\mathrm{ce}(\dxmaxm) = b + \frac{c-b}{1+e^{-d \dxmaxm}} \, .
\end{equation}
The optimal parameters are listed in \cite{geoceLDFgithub} and in \ref{sec:parametrizations}. We provide a video of the development of the charge-excess function with \dxmax as supplemental material \cite{videogeoce}.

\subsection{Extrapolation to larger zenith angles}

In this analysis we considered only air showers with zenith angles up to 60$^\circ$. This is because an additional asymmetry becomes relevant for larger zenith angles due to the projection onto the ground -- which we did not take into account in this work. 
The reason for this asymmetry is that at different observer positions the radio signal traversed different amounts of atmosphere until it reaches the ground. Observer positions 'below' the shower axis see smaller distances than positions 'above' the shower axis resulting in a left-right asymmetry which becomes relevant above 60$^\circ$ zenith angle. 

However, our results indicate that the parameters of the function increase monotonically with increasing \dxmax and we assured ourselves that our parametrizations follow the observed trend to at least $\dxmaxm = \SI{2000}{g/cm^2}$ (the corresponding zenith angles can be read off from Fig.~\ref{fig:starpattern} left). 
In particular for the observation altitude of \SI{1564}{m\, a.s.l.}, the extrapolation to larger \dxmax are similar to the \SI{0}{m\, a.s.l.} simulations that have $\sim$\SI{400}{g/cm^2} larger \dxmax values at 60$^\circ$ zenith angle (cf. Fig.~\ref{fig:starpattern} left and Fig.~\ref{fig:dxmax}).
Hence, our results can likely also be used at larger zenith angles (up to $\dxmaxm = \SI{2000}{g/cm^2}$) if the additional asymmetry due to the projection effect is taken into account. However, to make this model usable for horizontal air showers in general, the parametrization of the parameters of our function with \dxmax should be extended to larger \dxmax values using new CoREAS simulations.

\section{Combination to two-dimensional function}
With the results of the last two sections, we are able to describe the geomagnetic and charge-excess energy fluence distributions with only three parameters: The radiation energies of the two emission processes and the distance to the shower maximum. We can reduce the number of parameters further by noting that $\Ece$ can be expressed as a function of $\Egeo$ using the result of \cite{GlaserErad2016}: The relative charge-excess strength is a function of the air density at the shower maximum. With a model of the atmosphere, i.e., a description of the density as a function of height, the density at \xmax can be calculated from the distance to \xmax. Then, we can express both functions as a function of the complete radiation energy \Erad$ = \Egeo + \Ece$ and \dxmax. Then, the geomagnetic function reads
\begin{equation}
 f_\mathrm{geo} = 
\begin{cases}         
\frac{1}{N_{R_-} \times c_\mathrm{geo}} \frac{E_\mathrm{rad}}{1 + (a/\sin\alpha)^2} \, \, \exp\left(-\left(\frac{r-R_\mathrm{geo}}{\sqrt{2}\sigma_\mathrm{geo}}\right)^{p(r)} \right) &\mbox{if } R_\mathrm{geo} < 0 \\
\frac{1}{N_{R_+} \times c_\mathrm{geo}}  \frac{E_\mathrm{rad}}{1 + (a/\sin\alpha)^2} \left[\exp\left(-\left(\frac{r-R_\mathrm{geo}}{\sqrt{2}\sigma_\mathrm{geo}}\right)^{p(r)} \right) + \exp\left(-\left(\frac{r+R_\mathrm{geo}}{\sqrt{2}\sigma_\mathrm{geo}}\right)^{p(r)} \right)\right] &\mbox{if } R_\mathrm{geo} \geq 0 
\end{cases} \, ,
\label{eq:LDF_geo2}
\end{equation}
where $c_\mathrm{geo}$, $a$, \Rgeo, \sgeo and $p(r)$ are functions of only \dxmax and $c_\mathrm{geo}(\dxmaxm)$ is the parametrization of the ratio between $\Egeo$ and \Ageo. The parameter $a$ is the relative charge-excess strength and $\sin\alpha$ is the angle between the shower axis and the geomagnetic field. 
Similarly, the charge-excess function reads
\begin{equation}
 f_\mathrm{ce}(r) = \frac{1}{N_\mathrm{ce} \times c_\mathrm{ce}} \, E_\mathrm{rad} \left(1-\frac{1}{1 + (a/\sin\alpha)^2}\right) \, r^k \exp\left(\frac{-r^{p(r)} (k+1)}{p(r) \sigma_\mathrm{ce}^{p(r)}}\right) \, ,
 \label{eq:ce2}
\end{equation}
where $c_\mathrm{ce}$, $a$, $\sigma_\mathrm{ce}$, $k$ and $p(r)$ are functions of only \dxmax and $c_\mathrm{ce}(\dxmaxm)$ is the parametrization of the ratio between $\Ece$ and \Ace.

To obtain the total energy fluence at any position we need to combine the radially symmetric geomagnetic and charge-excess functions and take the interference between the two components into account. For the electric field, the simple relation 
\begin{equation}
 \vec{E} = \vec{E}_\mathrm{geo} + \vec{E}_\mathrm{ce} 
 \label{eq:Esum}
\end{equation}
holds at any position. We can write down this relation explicitly for the two components of $\vec{E}$:
\begin{align}
 \EvB(\vec{r}, t) =& E_\mathrm{geo}(\vec{r}, t) + \cos \phi \, E_\mathrm{ce}(\vec{r}, t) \\
 \EvvB(\vec{r}, t) =& \sin \phi \, E_\mathrm{ce}(\vec{r}, t)  \, ,
\end{align}
where $\phi = \arctantwo(y, x)$ is the polar angle in the \vBvvB plane as defined in Fig.~\ref{fig:starpattern} right and describes the position relative to the shower axis. From this relation and Eqs.~\eqref{eq:energyfluence1} + \eqref{eq:energyfluence2}, the interference in units of the energy fluence can be calculated \cite{GlaserErad2016}:
\begin{align}
 \fvB(\vec{r}) =& \left(\sqrt{\fgeo(r)} +  \cos \phi \sqrt{\fce(r)}\right)^2  \label{eq:fcomb1} \\
 \fvvB(\vec{r}) =& \sin^2 \phi \fce(r)  \label{eq:fcomb2} \\
 f =& \fvB + \fvvB
 \label{eq:fcomb3}
\end{align}

This calculation assumed that the geomagnetic and charge-excess component are in phase which introduces a small overestimation of $\fvB$ of 1\% as studied in \cite{GlaserErad2016}. In Sec.~\ref{sec:precision} we do not see that the resolution of the radiation energy and \dxmax is negatively impacted by this approximation and 1\% is anyway much smaller than the typical uncertainty on the energy fluence in an experiment of e.g. 5\% in case of the radio array of the Pierre Auger Observatory \cite{AERAEnergyPRD}. A phase difference can straightforwardly be introduced at the expense of an extra parameter by multiplying the $E_\mathrm{ce}$ component in Eq.~\eqref{eq:Esum} by a factor $\cos(\zeta)$ with $0\leq\zeta\leq\pi$ the relative phase difference
between the geomagnetic and charge excess component. Such a phase parameter could be included in a fit to experimental data.

\begin{figure}[t]
 \centering
 \includegraphics[width=1\textwidth]{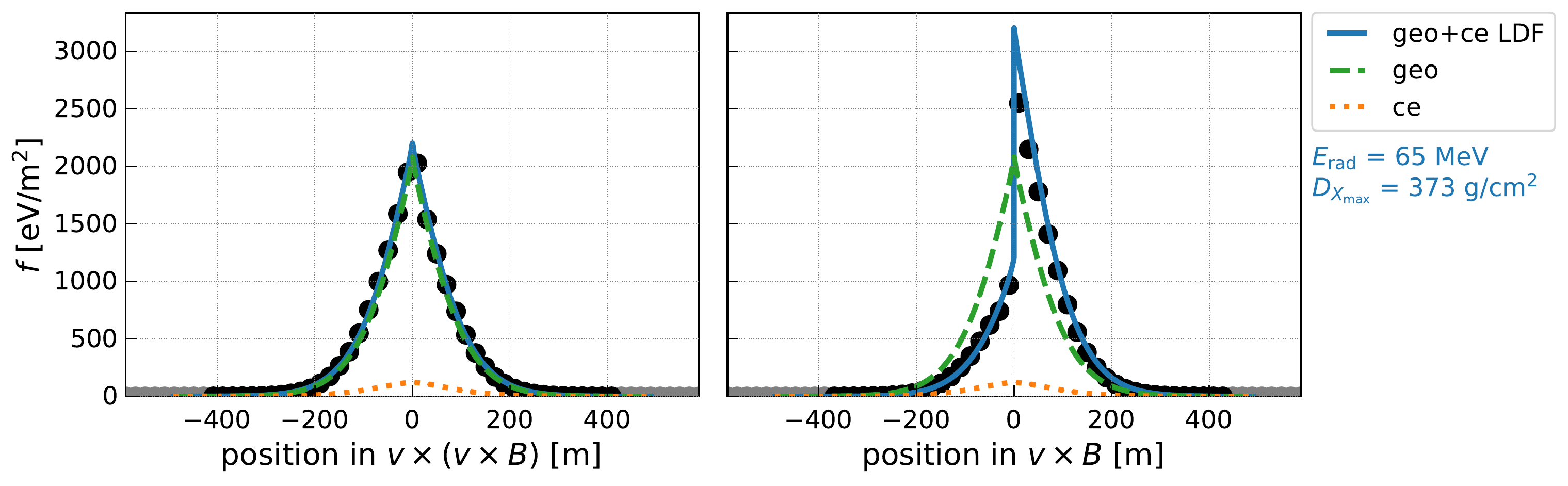}
 \caption{Same air shower as in Fig.~\ref{fig:example1}. The energy fluence is shown along the \vvB axis (no interference between geomagnetic and charge-excess) and along the \vB axis (maximum interference between geomagnetic and charge-excess). The signal distribution is modeled by the interplay between the geomagnetic function (green dashed curve) and the charge-excess function (orange dotted curve). The combined function is shown as a solid blue curve and depends only on \Erad and \dxmax. The interference is calculated via Eq.~\eqref{eq:fcomb1}, i.e., the square roots of the energy fluences of the two emission mechanisms are added and squared. Therefore, the small charge-excess component has a significant influence on the total signal.}
 \label{fig:example1_full}
\end{figure}

\begin{figure}[t]
 \centering
  \includegraphics[width=1\textwidth]{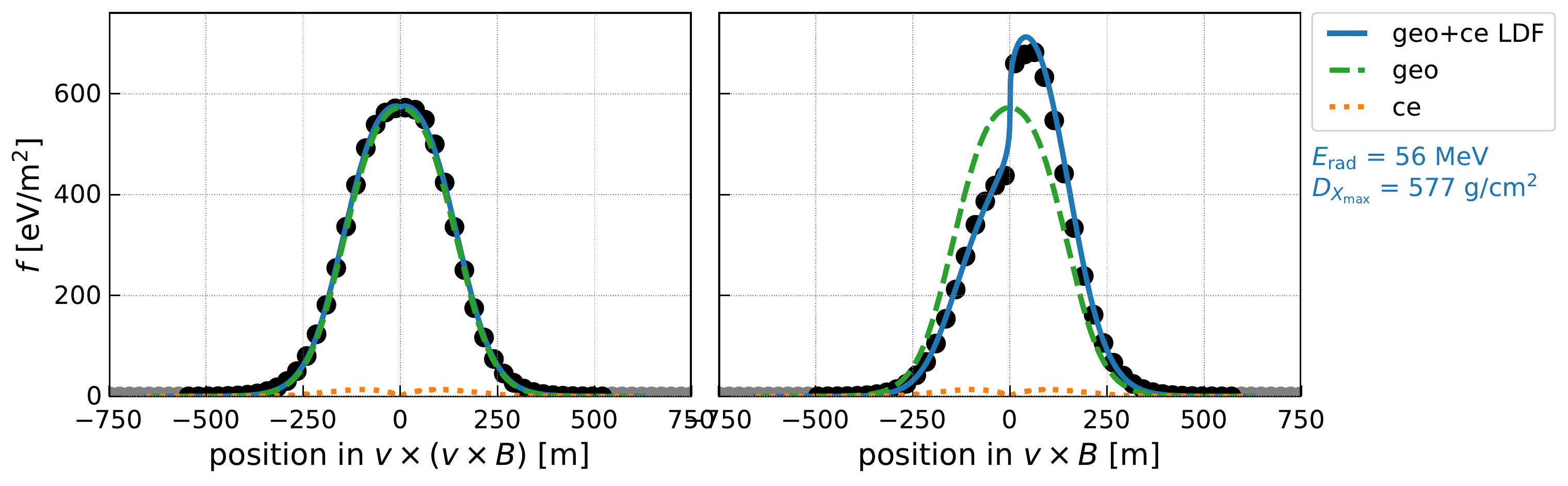}
  \caption{Same as Fig.~\ref{fig:example1_full} but for the air shower of Fig.~\ref{fig:example2}. }
  \label{fig:example2_full}
\end{figure}

\begin{figure}[t]
 \centering
  \includegraphics[width=1\textwidth]{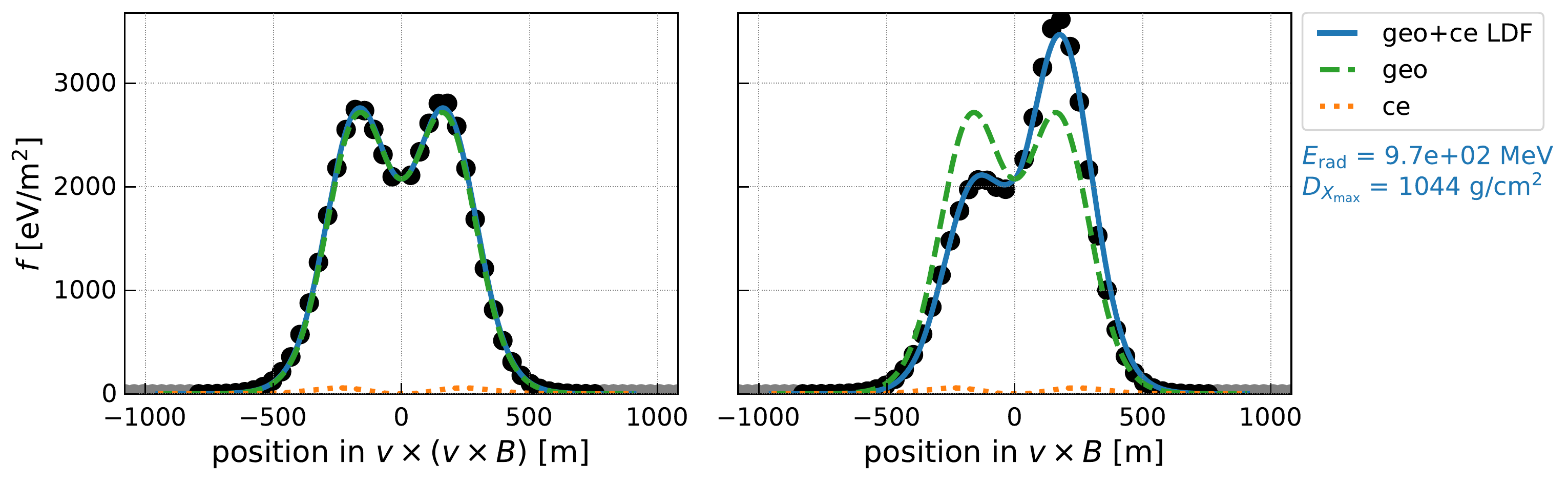}
  \caption{Same as Fig.~\ref{fig:example1_full} but for the air shower of Fig.~\ref{fig:example3}. } \label{fig:example3_full}
\end{figure}

Using these relations, the geomagnetic and charge-excess energy fluences can be combined to the total observed energy fluence at any position. In Figs.~\ref{fig:example1_full}-\ref{fig:example3_full}, the total energy fluence of our three example air showers at different distances to \xmax are presented with the optimal fit result of the combined two-dimensional function. We note that the function is completely defined by the two parameters $E_\mathrm{rad}$ and \dxmax, i.e., only these two parameters are optimized, and that the minimization is very stable once the function is parametrized to depend only on these two parameters. The energy fluence is shown along the \vvB axis in the left panels and along the \vB axis in right panels. Along the \vvB axis, the geomagnetic and charge-excess signals are polarized perpendicular to each other, hence we do not observe any interference between the two components. Along the \vB axis, the two components are polarized into the opposite direction for negative distances and are polarized into the same direction for positive distances. Hence, we observe a destructive  interference on one side of the shower axis and a constructive interference on the other side of the shower axis. This demonstrates that the observed asymmetry along the \vB axis is modeled well by the interference between geomagnetic and charge-excess emission. To better picture the evolution of the total observed energy fluence with \dxmax, we again provide a corresponding video as supplemental material \cite{videogeo}.

\subsection{Precision of analytic description}
\label{sec:precision}
In this section we quantify how well the function describes the simulated signal distribution. However, there is no unique way to do this. E.g., comparing the relative difference of energy fluences at each position will mostly highlight differences at large distances to the shower axis where the absolute difference is small. And calculating absolute differences of energy fluences normalized to the maximum of the function -- the method that was used in \cite{LOFARLDF} -- will mostly highlight (dis)agreement at the maximum energy fluence. Although describing the maxima of the distribution with high precision seems like the most important thing, it is not for most analyses. The physical quantities that are extracted from the distribution of the energy fluence is the radiation energy and \dxmax, and both of these quantities depend little on a precise modeling of the maximum amplitude. For determining the shape, it is more important to model the falling part of the distribution correctly. And for the radiation energy, the parts of the distribution where the product of energy fluence $f$ times the distance to the shower axis $r$ is largest are most important, because of the larger covered area. 

\begin{figure}[tbp]
 \centering
  \includegraphics[width=0.48\textwidth]{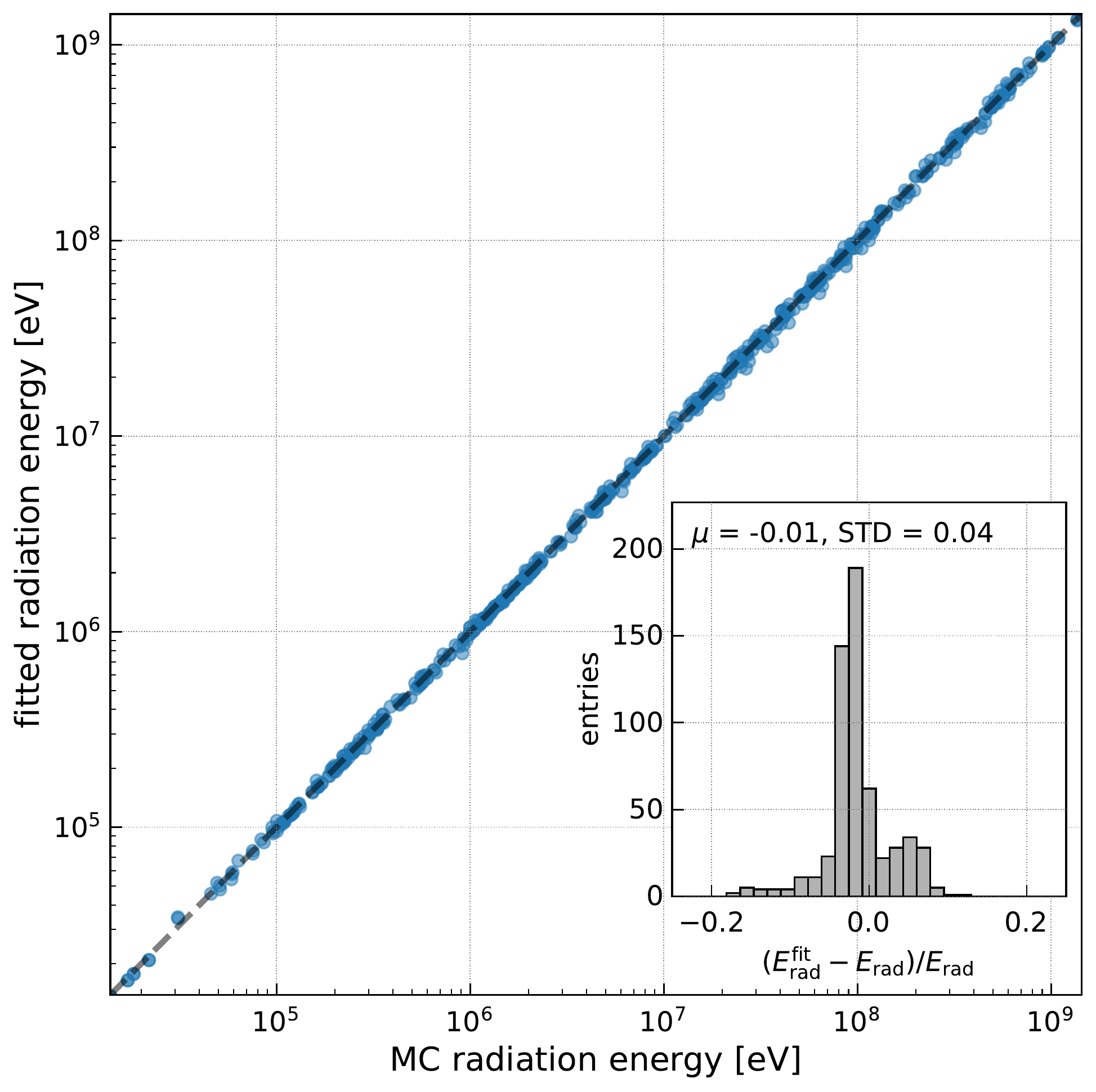}
  \includegraphics[width=0.48\textwidth]{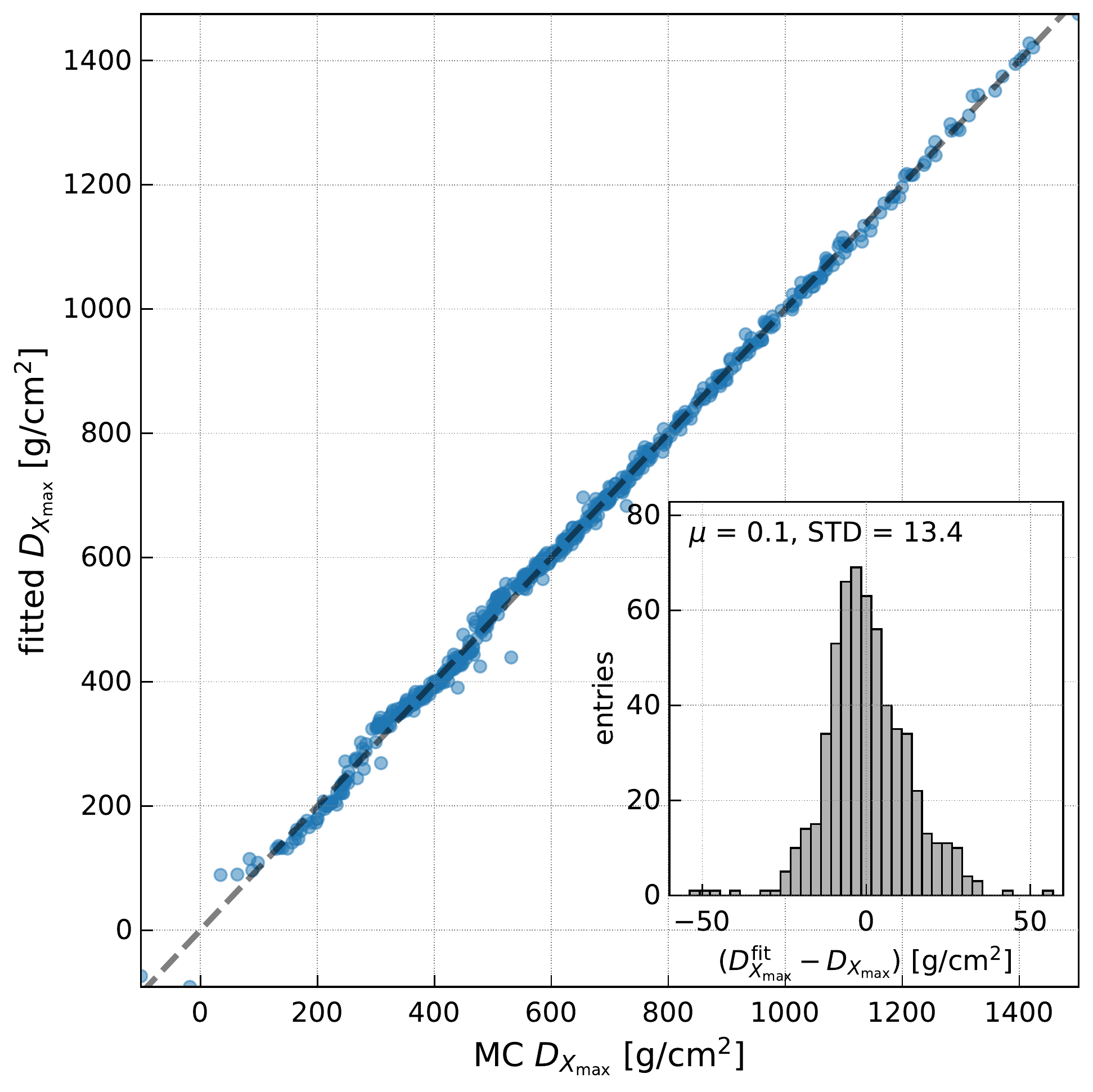}
  \caption{Performance of analytic description. The scatter plots show the optimal fit parameters \Erad and \dxmax versus the true MC values. The histograms show the respective relative deviation.}
  \label{fig:resolution}
\end{figure}

Therefore, we judge the precision of the analytic description by how well the radiation energy and \dxmax is extracted from the distribution of the energy fluence. This is done by using the parametrization that depends only on \Erad and \dxmax, i.e., Eqs.~\eqref{eq:fcomb1}-\eqref{eq:fcomb3} together with \eqref{eq:LDF_geo2} and \eqref{eq:ce2}. This function is fitted to the two-dimensional distribution of the energy fluence and the fitted values of the radiation energy and distance to \xmax are compared with the true MC values. The outcome of this study is presented in Fig.~\ref{fig:resolution}. The radiation energy can be determined with a resolution of 4\% and \dxmax with a resolution of \SI{13}{g/cm^2}. We note that with the knowledge of the zenith angle of the air shower \dxmax can be converted to \xmax, which is an estimator of the cosmic-ray mass. 
As these values are much smaller than the typical experimental uncertainties on \Erad and \dxmax that originate mostly from a finite sampling of the energy fluence and uncertainties in the measurement of the energy fluence itself, our analytic description is sufficiently good and does not limit the experimental resolution of sparse radio arrays. Even for detectors with a high station density, such as LOFAR, our model can serve as a fast alternative to the computationally intensive template matching technique \cite{LOFARNature2016} without dominating the \xmax resolution.

We developed this function with the prime goal of usability at radio arrays and shared our work with the Pierre Auger and the LOFAR collaboration right from the beginning. Therefore, this function has already been tested for sparse radio arrays and has even been successfully applied to data from the Pierre Auger Observatory and the LOFAR cosmic-ray detector \cite{ARENACanfora, ARENAPlaisier} to determine cosmic-ray properties.

\subsection{Usage at different observation altitudes and atmospheric conditions}
\label{sec:atm}
Some of the parameters of our function do not only depend on \dxmax but also show a slight dependence on the observation altitude as presented in Fig.~\ref{fig:dxmax}. Therefore, we presented separate parameterizations of \Rgeo, \sgeo and \sce for an observer at sea level and at \SI{1564}{m\, a.s.l.}. As the shapes of the parameterizations for the two observation altitudes are very similar, a pragmatic way to use this function for intermediate observation heights is to linearly interpolate between the two parameterizations. Inspecting Fig.~\ref{fig:dxmax} also allows to estimate an upper limit on the uncertainty for a specific observation height. To get a better assessment of the uncertainties, the fit results of this adjusted model can be compared to a few CoREAS simulations produced for the new observation altitude. 

This analysis was performed for a specific profile of the atmosphere that corresponds to the yearly average at the Pierre Auger Observatory. Different atmospheric conditions result in a change of \dxmax, e.g., the \dxmax values increase by \SI{10}{g/cm^2} for small \dxmax and up to \SI{20}{g/cm^2} for large \dxmax values if the atmospheric profile is changed to the US standard atmospheric model. As a consequence the \dxmax parameter of our function (called \dxmaxfit in the following) does not correspond to the true value \dxmaxtrue anymore. 
We note that the function still describes the data well and that the inferred radiation energy is accurate, it is just that the fit parameter \dxmaxfit will differ from the true \dxmaxtrue value. Hence, depending on the accuracy that is required, one can easily calculate the relation $D_{X_\mathrm{max}}^\mathrm{mymodel}$($D_{X_\mathrm{max}}^\mathrm{this model}$) \cite{radiotools} using only the two atmospheric models. To achieve a better accuracy, new CoREAS simulations for a particular atmospheric model need to be produced. Then, our model, as it is, can be fit to the new simulations to determine the relation $D_{X_\mathrm{max}}^\mathrm{new \, MC}$($D_{X_\mathrm{max}}^\mathrm{fit}$). With this prescription, different experiments do not need to go through the exercise of reparametrizing the 'low level' fit parameters (\sgeo, \Rgeo, \sce, $k$) with \dxmax, and the number of required simulations will be less than the ones used in this work.

\section{Conclusion}
Ultra-high energy cosmic rays can be measured through the detection of radio-frequency radiation from air showers. The radio emission originates from deflections of the air-shower particles in the geomagnetic field and from a time-varying negative charge excess in the shower front. The distribution of the radio signal on the ground contains information on crucial cosmic-ray properties, such as energy and mass. The strength of the radio emission scales with the cosmic-ray energy and the shape of the spatial signal distribution depends on the distance to the emission region. A long standing challenge to access this information experimentally with a sparse grid of antennas is a corresponding analytic description.

We have presented a new analytic model of the radio signal distribution that models the spatial distribution of the energy fluence originating from the two emission mechanism separately. The observed two-dimensional asymmetry in the signal distribution is modeled by the interference between the two emission mechanisms. 
Thereby, we explicitly take into account the polarization of the radio signal by separately describing the \vB and \vvB component of the energy fluence. Hence, the available information at a single antenna station is doubled which allows for a more precise determination of the signal distribution compared to previous models for the same number of detector stations. 

One parameter of our model is the radiation energy, which directly relates to the electromagnetic shower energy. The contribution to the resolution of the radiation energy of our model is 4\% which translates to an uncertainty of the cosmic-ray energy of only 2\% due to a quadratic scaling between the two quantities. This is negligible to practical sampling uncertainties of real radio detector arrays. Hence, our model is particularly useful to precisely determine the cosmic-ray energy from radio air-shower measurements. 

The remaining parameters of our function are correlated to the distance to the shower maximum. We have shown that this model can determine the distance to shower maximum to a precision of \SI{13}{g/cm^2}. The experimental sampling limitation of a real radio array will likely dominate the final uncertainty also here. In addition, there will be some deterioration when converting \dxmax into \xmax, especially
at larger zenith angle, due to the zenith angle resolution of the shower and atmospheric conditions uncertainties. However, the fitting procedure presented here is unlikely to be the limiting factor in the final \xmax resolution for practical radio arrays.
Hence, we can formulate our model to depend only on the radiation energy, the distance to \xmax and the core position, which was always at the coordinate origin in this simulation study. Thus, our model provides direct access to the main air-shower parameters energy and \xmax and manages to use the smallest possible number of parameters. 

Furthermore, our model provides an absolute prediction of the energy fluence at any position for a given air-shower energy and \xmax. For many studies our model can replace computational-extensive simulation studies, e.g., for estimating the sensitivity of a detector.
We provide a reference implementation of our model in \emph{Python} \cite{geoceLDFgithub}.

\section*{Acknowledgements}

We would like to thank Hans Dembinski for making us aware of the advantages of Spline functions and Fabrizia Canfora for performing several cross-checks using different atmospheric models. We furthermore acknowledge fruitful discussions with our colleagues in the AERA task of the Pierre Auger Collaboration.
This work is supported by the Ministry of Innovation, Science and Research of the State of North
Rhine-Westphalia, the Federal Ministry of Education and Research (BMBF), the Deutsche Forschungsgesellschaft (grant GL 914/1-1), and the Netherlands Organisation for Scientific Research (NWO).

\appendix
\section{Normalization}
\label{sec:normalization}
The normalizations of the geomagnetic and charge-excess functions Eqs.~\eqref{eq:LDF_geo} and \eqref{eq:ce} are given by 
\begin{equation}
 N_{R_-} = \sigma \pi \sqrt{2} \left[\sqrt{\pi} R \, \mathrm{erfc}\left(\frac{-R}{\sqrt{2}\sigma}\right)+ \sqrt{2} \sigma \exp\left(\frac{-R^2}{2\sigma^2}\right)\right]
\end{equation}
\begin{equation}
 N_{R_+} = 2 \pi \sigma \left[\mathrm{erf}\left(\frac{R}{\sqrt{2}\sigma}\right) \sqrt{2\pi} R + 2 \sigma \exp\left(\frac{-R^2}{2\sigma^2}\right) \right]
\end{equation}
\begin{equation}
 N_{ce} = \frac{ 2 \pi}{k + 1} 2^k (2k + 2)^{-0.5k} \, \sigma^{k+2} \, \Gamma(k/2 + 1) \, .
\end{equation} 

\section{Spline functions}
\label{sec:spline}
Spline functions are piece wise polynomial functions. They are defined by an array of knots, the places where the pieces meet, and a corresponding set of coefficients. An important property of splines is that they are continuous at the knots and, hence, can be used to obtain a smooth parametrization. 

We use cubic B-splines and determine the required number of knots and the optimal coefficients of the splines in a minimization using the \emph{UnivariateSpline} method of the \emph{scipy interpolate} python package. The following function is minimized
\begin{equation}
\sum_i [y_i - S(x_i)]^2 < s \, ,
\end{equation}
where $S(x)$ is the spline function. The accuracy of the interpolation is adjusted by specifying a smoothing condition $s$. The number of knots will be increased until the smoothing condition is satisfied. 

\section{Parametrizations of distance to Xmax dependencies}
\label{sec:parametrizations}

For the charge-excess function the dependence of $b$ on \dxmax is described by
\[ b(\dxmaxm)  = \begin{cases} 
      147 - \SI{0.251}{cm^2/g}  \dxmaxm & k < 10^{-5} \\
      56 + \SI{0.324}{cm^2/g}  \dxmaxm & k \geq 10^{-5}
   \end{cases}
\]

For the charge-excess function the dependence of $r_\mathrm{cut}$ on \dxmax is described by
\[ r_\mathrm{cut}(\dxmaxm)  =  \begin{cases} 
      0 & k < 10^{-5} \\
      \frac{-p_1 + \sqrt{(4  b(\dxmaxm) - 4  p_0)  p_2 + p_1 ^ 2}}{2 p_2} & k \geq   10^{-5} 
   \end{cases}
\]
with $p_0 = 29.057$, $p_1 = \SI{0.197}{1/m}$ and $p_2 = \SI{1.80589e-3}{1/m^2}$.

For the charge-excess function the dependence of $k$ on \dxmax is described by
\begin{equation}
 k(\dxmaxm) = \max\left(0, b + \frac{c - b}{1 + \exp(-d  (\dxmaxm - a))}\right)   \, ,
\end{equation}
with $a=\SI{5.80505613e+02}{g/cm^2}, b=-1.76588481, c=3.12029983, d=\SI{3.73038601e-03}{cm^2/g}$.

The parameters that define the various spline functions are presented in the reference implementation \cite{geoceLDFgithub} and in the supplemental material \cite{splineconstants}.


\section*{References}

\begin{thebibliography}{10}
\expandafter\ifx\csname url\endcsname\relax
  \def\url#1{\texttt{#1}}\fi
\expandafter\ifx\csname urlprefix\endcsname\relax\def\urlprefix{URL }\fi
\expandafter\ifx\csname href\endcsname\relax
  \def\href#1#2{#2} \def\path#1{#1}\fi

\bibitem{Auger2014}
A.~Aab et al., \href{http://dx.doi.org/10.1016/j.nima.2015.06.058}{{The Pierre
  Auger Cosmic Ray Observatory}}, Nucl. Instrum. Meth. A 798 (2015) 172--213.
\newblock \href {http://dx.doi.org/10.1016/j.nima.2015.06.058}
  {\path{doi:10.1016/j.nima.2015.06.058}}.

\bibitem{TA2008}
H.~Kawai et al.,
  \href{http://linkinghub.elsevier.com/retrieve/pii/S0920563207007992}{{Telescope
  Array Experiment}}, Nucl. Phys. B, Proc. Suppl. 175-176 (2008) 221--226.
\newblock \href {http://dx.doi.org/10.1016/j.nuclphysbps.2007.11.002}
  {\path{doi:10.1016/j.nuclphysbps.2007.11.002}}.

\bibitem{TunkaRex2015}
P.~Bezyazeekov et al.,
  \href{http://www.sciencedirect.com/science/article/pii/S0168900215010256}{{Measurement
  of cosmic-ray air showers with the {Tunka} Radio Extension
  {(Tunka-Rex)}}}, Nucl. Instrum. Meth. A 802 (2015) 89--96.
\newblock \href {http://dx.doi.org/10.1016/j.nima.2015.08.061}
  {\path{doi:10.1016/j.nima.2015.08.061}}.


\bibitem{Huege2016}
T.~Huege, \href{http://dx.doi.org/10.1016/j.physrep.2016.02.001}{{Radio
  detection of cosmic ray air showers in the digital era}}, Phys. Rep. 620
  (2016) 1--52.
\newblock \href {http://dx.doi.org/10.1016/j.physrep.2016.02.001}
  {\path{doi:10.1016/j.physrep.2016.02.001}}.

\bibitem{Schroeder2016}
F.~G. Schr{\"{o}}der, {Radio detection of cosmic-ray air showers and
  high-energy neutrinos}, Prog. Part. Nucl. Phys. 93 (2017) 1--68.
\newblock \href {http://dx.doi.org/10.1016/j.ppnp.2016.12.002}
  {\path{doi:10.1016/j.ppnp.2016.12.002}}.

\bibitem{LOFAREnergy}
A.~Nelles et al., \href{http://stacks.iop.org/1475-7516/2015/i=05/a=018}{{The
  radio emission pattern of air showers as measured with {LOFAR} -- a
  tool for the reconstruction of the energy and the shower maximum}}, J.
  Cosmol. Astropart. Phys. 05(2015)18.
\newblock \href {http://dx.doi.org/10.1088/1475-7516/2015/05/018}
  {\path{doi:10.1088/1475-7516/2015/05/018}}.

\bibitem{AERAEnergyPRL}
A.~Aab et al.,
  \href{http://dx.doi.org/10.1103/PhysRevLett.116.241101}{{Measurement of the
  Radiation Energy in the Radio Signal of Extensive Air Showers as a Universal
  Estimator of Cosmic-Ray Energy}}, Phys. Rev. Lett. 116 (2016) 241101.
\newblock \href {http://dx.doi.org/10.1103/PhysRevLett.116.241101}
  {\path{doi:10.1103/PhysRevLett.116.241101}}.

\bibitem{AERAEnergyPRD}
A.~Aab et al., \href{http://dx.doi.org/10.1103/PhysRevD.93.122005}{{Energy
  Estimation of Cosmic Rays with the {Engineering Radio Array} of the
  {Pierre Auger Observatory}}}, Phys. Rev. D 93 (2016) 122005.
\newblock \href {http://dx.doi.org/10.1103/PhysRevD.93.122005}
  {\path{doi:10.1103/PhysRevD.93.122005}}.

\bibitem{LOFARNature2016}
S.~Buitink et al., \href{http://dx.doi.org/10.1038/nature16976}{{A large
  light-mass component of cosmic rays at {$10^{17} -
  10^{17.5}$} electronvolts from radio observations}}, Nature
  531~(7592) (2016) 70--73.
\newblock \href {http://dx.doi.org/10.1038/nature16976}
  {\path{doi:10.1038/nature16976}}.

\bibitem{PhDGlaser}
C.~Glaser, {Absolute energy calibration of the {Pierre Auger Observatory}
  using radio emission of extensive air showers}, Ph.D. thesis, RWTH Aachen
  University (2017).
\newblock \href {http://dx.doi.org/10.18154/RWTH-2017-02960}
  {\path{doi:10.18154/RWTH-2017-02960}}.

\bibitem{KrauseICRC2017}
{R. Krause for the Pierre Auger Collaboration}, {A new method to determine the
  energy scale for high-energy cosmic rays using radio measurements at the
  Pierre Auger Observatory}, PoS(ICRC2017)528.

\bibitem{GlaserErad2016}
C.~Glaser, M.~Erdmann, J.~R. H{\"{o}}randel, T.~Huege, J.~Schulz, {Simulation
  of Radiation Energy Release in Air Showers}, J. Cosmol. Astropart. Phys.
  09 (2016) 24.
\newblock \href {http://dx.doi.org/10.1088/1475-7516/2016/09/024}
  {\path{doi:10.1088/1475-7516/2016/09/024}}.

\bibitem{Gottowik2017}
M.~Gottowik, C.~Glaser, T.~Huege, J.~Rautenberg,
  \href{https://doi.org/10.1016/j.astropartphys.2018.07.004}{{Determination of the absolute energy
  scale of extensive air showers via radio emission: systematic uncertainty of
  underlying first-principle calculations}}, Astropart. Phys. 103 (2018) 87--93

\bibitem{Endpoint2011}
C.~W. James, H.~Falcke, T.~Huege, M.~Ludwig, {General description of
  electromagnetic radiation processes based on instantaneous charge
  acceleration in endpoints}, Phys. Rev. E 84 (2011) 56602.
\newblock \href {http://dx.doi.org/10.1103/physreve.84.056602}
  {\path{doi:10.1103/physreve.84.056602}}.

\bibitem{Allan1971a}
H.~Allan, \href{http://adsabs.harvard.edu/abs/1971ICRC....3.1108A}{{The Lateral
  Distribution of the Radio Emission, and its Dependence on the Longitudinal
  Structure of the Air Shower}}, Proc. 12th Int. Conf. Cosm. Rays 3 (1971)
  1108.


\bibitem{LOFARLDF}
A.~Nelles et al.,
  \href{http://linkinghub.elsevier.com/retrieve/pii/S0927650514000565}{{A
  parameterization for the radio emission of air showers as predicted by CoREAS
  simulations and applied to LOFAR measurements}}, Astropart. Phys. 60 (2015)
  13--24.
\newblock \href {http://dx.doi.org/10.1016/j.astropartphys.2014.05.001}
  {\path{doi:10.1016/j.astropartphys.2014.05.001}}.


\bibitem{Alvarez-Muniz2014a}
J.~Alvarez-Mu{\~{n}}iz, W.~R. Carvalho, H.~Schoorlemmer, E.~Zas, {Radio pulses
  from ultra-high energy atmospheric showers as the superposition of
  {Askaryan} and geomagnetic mechanisms}, Astropart. Phys. 59 (2014)
  29--38.
\newblock \href {http://dx.doi.org/10.1016/j.astropartphys.2014.04.004}
  {\path{doi:10.1016/j.astropartphys.2014.04.004}}.

\bibitem{geoceLDFgithub}
\href{https://github.com/cg-laser/geoceLDF}{{GeoCELDF Reference
  Implementation}}.
\newline\urlprefix\url{https://github.com/cg-laser/geoceLDF}

\bibitem{CoREAS2013}
T.~Huege, M.~Ludwig, C.~W. James, {Simulating radio emission from air showers
  with {CoREAS}}, AIP Conf. Proc. 1535 (2013) 128--132.
\newblock \href {http://dx.doi.org/10.1063/1.4807534}
  {\path{doi:10.1063/1.4807534}}.

\bibitem{Corsika}
D.~Heck, G.~Schatz, T.~Thouw, J.~Knapp, J.~N. Capdevielle, {CORSIKA: A
  {Monte Carlo} code to simulate extensive air showers}, Rep. FZKA 6019.

\bibitem{LofarXmaxMethod}
S.~Buitink et al., \href{http://arxiv.org/pdf/1408.7001.pdf
  http://arxiv.org/abs/1408.7001
  http://link.aps.org/doi/10.1103/PhysRevD.90.082003}{{Method for high
  precision reconstruction of air shower Xmax using two-dimensional radio
  intensity profiles}}, Phys. Rev. D 90 (2014) 82003.
\newblock \href {http://dx.doi.org/10.1103/PhysRevD.90.082003}
  {\path{doi:10.1103/PhysRevD.90.082003}}.


\bibitem{ZHAireS2012a}
J.~Alvarez-Mu{\~{n}}iz, W.~R. Carvalho, M.~Tueros, E.~Zas,
  \href{http://dx.doi.org/10.1016/j.astropartphys.2011.10.002}{{Coherent
  Cherenkov radio pulses from hadronic showers up to {EeV} energies}},
  Astropart. Phys. 35 (2012) 287--299.
\newblock \href {http://dx.doi.org/10.1016/j.astropartphys.2011.10.002}
  {\path{doi:10.1016/j.astropartphys.2011.10.002}}.


\bibitem{QGSJet}
S.~Ostapchenko, {{Monte} {Carlo} treatment of hadronic interactions
  in enhanced Pomeron scheme: {QGSJET}-{II} model} 83 (2011)
  14018.
\newblock \href {http://dx.doi.org/10.1103/physrevd.83.014018}
  {\path{doi:10.1103/physrevd.83.014018}}.

\bibitem{URQMD}
M.~Bleicher, E.~Zabrodin, C.~Spieles, S.~A. Bass, C.~Ernst, S.~Soff,
  L.~Bravina, M.~Belkacem, H.~Weber, H.~St{\"{o}}cker, W.~Greiner,
  {Relativistic hadron-hadron collisions in the ultra-relativistic quantum
  molecular dynamics model} 25 (1999) 1859--1896.

\bibitem{videogeoce}
{Supplemental material: Video showing the evolution of the geomagnetic and
  charg-excess function with distance to Xmax \href{https://www.sciencedirect.com/science/article/pii/S0927650518301579?via%3Dihub#sec0023}{doi:10.1016/j.astropartphys.2018.08.004}}.

\bibitem{videogeo}
{Supplemental material: Video showing the evolution of the energy fluence with
  distance to Xmax \href{https://www.sciencedirect.com/science/article/pii/S0927650518301579?via%3Dihub#sec0023}{doi:10.1016/j.astropartphys.2018.08.004}}.

  
  
\bibitem{ARENACanfora}
{F.~Canfora for the Pierre Auger Collaboration}, \href{https://indico.cern.ch/event/667036/contributions/3002762/}{{Cosmic ray composition measurements with the Auger Engineering Radio Array}}, 8th International Conference on Acoustic and Radio EeV Neutrino Detection Activities (ARENA2018), Catania, Italy June 2018.

\bibitem{ARENAPlaisier}
{I.~Plaisier et al.}, \href{https://indico.cern.ch/event/667036/contributions/3003418/}{{A new LDF parametrization for the air shower radio footprint applied to LOFAR data}}, 8th International Conference on Acoustic and Radio EeV Neutrino Detection Activities (ARENA2018), Catania, Italy June 2018.

\bibitem{radiotools}
  {To do so one could use the atmosphere class of the \emph{radiotools} package that implements the relevant functions and most atmophere models. \url{https://github.com/cg-laser/radiotools/blob/master/radiotools/atmosphere/models.py}}

\bibitem{splineconstants}
{Supplemental material: Parameters of the spline functions \href{https://www.sciencedirect.com/science/article/pii/S0927650518301579?via%3Dihub#sec0023}{doi:10.1016/j.astropartphys.2018.08.004}}.



\end{thebibliography}
\end{document}